\documentclass[a4paper,11pt]{article}
 \pdfoutput=1
\usepackage{jheppub}
\usepackage[caption=false,labelfont={sf,bf}, textfont=sf,hangindent=18pt,justification=raggedright]{subfig}
\usepackage{ulem}

\newcommand{\dfourx}{\mathrm{d}^4 x}
\newcommand{\dx}{\mathrm{d}x}
\newcommand{\dt}{\mathrm{d}t}
\renewcommand{\Im}{\mathrm{Im}}
\newcommand{\as}{\alpha_s}
\newcommand{\borel}{\hat{\mathcal{B}}}
\newcommand{\quarkthreed}{\langle\overline{q}q\rangle}
\newcommand{\gluonfourd}{\langle\alpha G^2\rangle}
\newcommand{\mixed}{\langle g\overline{q} \sigma G q\rangle}
\newcommand{\gluonsixd}{\langle g^3 G^3\rangle}
\newcommand{\vev}[1]{\langle\Omega|#1|\Omega\rangle}
\newcommand{\angled}[1]{\langle #1\rangle}
\newcommand{\dilog}{\mathrm{Li}_2}
\newcommand{\lsr}{\mathcal{R}}
\newcommand{\gev}{\ensuremath{\text{GeV}}}
\newcommand{\mev}{\ensuremath{\text{MeV}}}
\newcommand{\msbar}{$\overline{\text{MS}}$}
\title{Masses of Open-Flavour Heavy-Light Hybrids from QCD Sum-Rules}
\author[a,1]{J.~Ho,\note{Corresponding author.}}
\author[b]{D.~Harnett,}
\author[a]{T.G.~Steele}

\affiliation[a]{Department of Physics and Engineering Physics,\\ University of Saskatchewan,\\           Saskatoon, SK, S7N 5E2, Canada}
\affiliation[b]{Department of Physics,\\ University of the Fraser Valley,\\ Abbotsford, BC, V2S 7M8, Canada}

\emailAdd{j.ho@usask.ca}
\emailAdd{derek.harnett@ufv.ca}
\emailAdd{tom.steele@usask.ca}

\arxivnumber{1609.06750}

\abstract{We use QCD Laplace sum-rules to predict masses of open-flavour heavy-light hybrids where
one of the hybrid's constituent quarks is a charm or bottom 
and the other is an up, down, or strange.
We compute leading-order, diagonal correlation functions of several hybrid interpolating currents, 
taking into account QCD condensates up to dimension-six, and 
extract hybrid mass predictions for all $J^P\in\{0^{\pm},\,1^{\pm}\}$, as well as explore possible mixing effects with conventional quark-antiquark mesons.
Within theoretical uncertainties, our results are consistent with a degeneracy between the heavy-nonstrange and heavy-strange 
hybrids in all $J^P$ channels.  
We find a similar mass hierarchy of $1^+$, $1^{-}$, and $0^+$ states (a $1^{+}$ state lighter than essentially degenerate $1^{-}$ and $0^{+}$ states) in both the charm and bottom sectors, and discuss an interpretation for the $0^-$ states.
If conventional meson mixing is present the effect is an increase in the hybrid mass prediction, and we estimate an upper bound on this effect.
}
\begin{document}
\maketitle
\flushbottom
\section{Introduction}\label{I}

\noindent Hybrids are hypothesized, beyond-the-quark-model hadrons that exhibit explicit 
quark, antiquark, and gluonic degrees of freedom.
They are colour singlets and so should be permissible within quantum 
chromodynamics (QCD);
thus, the question of their existence provides us with a
key test of our characterization of confinement.
Despite nearly four decades of searching, hybrids have not yet been conclusively identified in experiment.
There are, however, a number of noteworthy candidates.
For example, the Particle Data Group (PDG)~\cite{OliveEtAl2014} lists a pair of tentative 
resonances, the $\pi_1(1400)$ and the $\pi_1(1600)$, both with exotic $J^{PC}=1^{-+}$, 
a combination inaccessible to conventional quark-antiquark mesons \cite{Narison2009,Meyer2015}.
There are several non-exotic hybrid prospects as well.
For instance, each of the resonances $\phi(2170)$, X(3872), Y(3940), and Y(4260)
has been singled out as a possible hybrid or at least as a mixed hadron 
containing a hybrid component \cite{DingYan2007,ChenJinKleivSteeleWangXu2013,Zhu2005,Belle2005,BergHarnettKleivSteele2012,Ketzer2012}.

Definitively assigning a hybrid interpretation to an observed resonance 
would be greatly facilitated by agreement between theory and experiment concerning 
the candidate hybrid's mass.
Previous calculations aimed at predicting hybrid masses have been made using
a constituent gluon model~\cite{HornMandula1978}, the MIT bag model~\cite{BarnesCloseVironEtAl1983,ChanowitzSharpe1983}, 
and the flux tube model~\cite{IsgurKokoskiPaton1985,ClosePage1995,BarnesCloseSwanson1995}
as well as through the QCD-based approaches of QCD sum-rules~\cite{Narison2009,BergHarnettKleivSteele2012,GovaertsVironGusbinEtAl1983,GovaertsVironGusbinEtAl1984,LatorreNarisonPascualEtAl1984,LatorrePascualNarison1987,BalitskyDiakonovYung1986,JinKoernerSteele2003,Huang2015,ChetyrkinNarison2000,Zhu097502,HarnettKleivSteeleJin2012,ChenKleivSteeleEtAl2013,QiaoTangHaoLi2012},
lattice QCD~\cite{PerantonisMichael1990,LiuMoirPeardonEtAl2012,MoirPeardonRyanEtAl2013,Dudek2011}, and Heavy Quark Effective Theory~\cite{HuangJinZhang2000}.
Unfortunately, as of yet, there is little consensus concerning hybrid masses.

To date, closed-flavour (hidden-flavour or quarkonium) hybrids have received 
more attention than open-flavour hybrids
likely because most promising hybrid candidates are closed.
Furthermore, closed-flavour hybrids allow for exotic $J^{PC}$ quantum numbers;
open-flavour hybrids, on the other hand, are not eigenstates of C-parity, and so
are characterized by non-exotic $J^P$ quantum numbers. 
However, the recent observation of the fully-open-flavour X(5568) 
containing a heavy (bottom) quark~\cite{D0Collab2016,LHCbCollab2016} 
may be a precursor to additional open-flavour discoveries that do not have a simple quark-model explanation
(e.g., the X(5568) has been studied as a $\bar b\bar d su$ 
tetraquark \cite{ChenChenLiuSteeleZhu2016}). 
Hence, computing masses of open hybrids containing heavy quarks is timely 
and of phenomenological relevance.

Ground state masses of bottom-charm hybrids 
were recently computed using QCD sum-rules in~\cite{ChenSteeleZhu2014}; therefore, we focus on a QCD sum-rules analysis of open-flavour heavy-light hybrids
i.e., hybrids containing one heavy quark (charm or bottom) and one light quark (up, down, or strange).

The seminal application of QCD Laplace sum-rules to open-flavour hybrids was performed by 
Govaerts, Reinders, and Weyers~\cite{GovaertsReindersWeyers1985} (hereafter referred to as GRW).
Therein, they considered four distinct currents covering $J\in\{0,\,1\}$
in an effort to compute a comprehensive collection of hybrid masses.
Their QCD correlator calculations took into account perturbation theory as well as 
mass-dimension-three (i.e., 3d) quark and 4d gluon condensate contributions. Precisely half of the analyses stabilized 
and yielded viable mass predictions.
However, for all heavy-light hybrids, the ground state hybrid mass was
uncomfortably close to the continuum threshold (with a typical separation
of roughly $10\,\mev$), so that even a modest hadron width
would result in the resonance essentially merging with the continuum ~\cite{GovaertsReindersWeyers1985}.

In this article, we extend the work of GRW by  
including both 5d mixed and 6d gluon condensate contributions in
our correlator calculations.
As noted in GRW, for open-flavour heavy-light hybrids, condensates involving light quarks
could be enhanced by a heavy quark mass allowing for the possibility
of a numerically significant contribution to the sum-rules.  
By this reasoning, the 5d mixed condensate should also be included.
As for the 6d gluon condensate, recent sum-rules analyses of closed-flavour 
heavy hybrids~\cite{BergHarnettKleivSteele2012,HarnettKleivSteeleJin2012,ChenKleivSteeleEtAl2013,QiaoTangHaoLi2012} have demonstrated that it is important 
and can have a stabilizing effect on what were, in the pioneering 
work~\cite{GovaertsReindersRubinsteinEtAl1985,GovaertsReindersFranckenEtAl1987}, 
unstable analyses. We also consider the possibility that conventional quark-antiquark mesons couple to the hybrid current, and demonstrate that this leads to an increase in the predicted value of the hybrid mass. A methodology is developed to estimate an upper bound on this mass increase in each channel.

This paper is organized as follows:
in Section~\ref{II}, we define the currents that we use to probe open-flavour heavy-light
hybrids and compute corresponding correlation functions;
in Section~\ref{III}, we 
generate QCD sum-rules for each of the correlators;
in Section~\ref{IV}, we present our analysis methodology
as well as our mass predictions for those channels which stabilized; in Section~\ref{new_section} we consider the effects of mixing; 
and, in Section~\ref{V}, we discuss our results 
and compare them to GRW
and to contemporary predictions made using lattice QCD.

\section{Currents and Correlators}\label{II}
Following GRW, we define open-flavour heavy-light hybrid interpolating currents
\begin{equation}\label{current}
  j_{\mu}=\frac{g_s}{2} \overline{Q} \Gamma^{\rho} \lambda^a q \mathcal{G}^a_{\mu\rho}
\end{equation}
where $g_s$ is the strong coupling and $\lambda^a$ are the Gell-Mann matrices.
The field $Q$ represents a heavy charm or bottom quark with mass $M_Q$
whereas $q$ represents a light up, down, or strange quark with mass $m_q$. 
The Dirac matrix $\Gamma^{\rho}$ satisfies
\begin{equation}
  \Gamma^{\rho} \in \{\gamma^{\rho},\,\gamma^{\rho}\gamma_5\},\label{Gamma_definition}\\
\end{equation}
and the tensor $\mathcal{G}^a_{\mu\rho}$ satisfies
\begin{equation}
  \mathcal{G}^a_{\mu\rho} \in \{G^a_{\mu\rho},\,\tilde{G}^a_{\mu\rho}\}\label{H_definition}
\end{equation}
where $G^a_{\mu\rho}$ is the gluon field strength and 
\begin{equation}
  \tilde{G}^a_{\mu\rho} = \frac{1}{2}\epsilon_{\mu\rho\nu\sigma}G^a_{\nu\sigma}
\end{equation}
is its dual defined using the totally antisymmetric Levi-Civita 
symbol $\epsilon_{\mu\rho\nu\sigma}$.

For each of the four currents defined through~(\ref{current})--(\ref{H_definition}), 
we consider a diagonal correlation function
\begin{align}
  \Pi_{\mu\nu}(q) &= i\int\dfourx \,e^{i q\cdot x} 
    \vev{\tau j_{\mu}(x)j^{\dag}_{\nu}(0)}\label{correlator}\\
  &= \frac{q_{\mu}q_{\nu}}{q^2}\Pi^{(0)}(q^2) 
   + \left(\frac{q_{\mu}q_{\nu}}{q^2}-g_{\mu\nu}\right)\Pi^{(1)}(q^2)\,,\label{spin_breakdown}
\end{align}
where $\Pi^{(0)}$ probes spin-0 states and $\Pi^{(1)}$ probes spin-1 states.
Each of $\Pi^{(0)}$ and $\Pi^{(1)}$ couples to a particular parity value, and,
in the case of closed-flavour hybrids, also to a particular C-parity value; 
however, as noted in Section~\ref{I}, open-flavour hybrids are not C-parity eigenstates.
Regardless, we will refer to $\Pi^{(0)}$ and $\Pi^{(1)}$
using the $J^{PC}$ assignments they would have if we were investigating closed- 
rather than open-flavour hybrids. 
But, to stress that the $C$-value cannot be taken 
literally, we will enclose it in brackets (a notation employed in~\cite{ChenSteeleZhu2014,HilgerKrassnigg}).
In Table~\ref{JPC_table}, we provide a breakdown of which currents couple to which 
$J^{P(C)}$ combinations.
\begin{table}[hb]
\centering
\caption{The $J^{P(C)}$ combinations probed through different choices of 
$\Gamma^{\rho}$~(\ref{Gamma_definition}) and $\mathcal{G}^a_{\mu\rho}$~(\ref{H_definition}).}
\label{JPC_table}
\begin{tabular}{c|c||c}
  $\Gamma^{\rho}$ & $\mathcal{G}^a_{\mu\rho}$ & $J^{P(C)}$\\
  \hline
  $\gamma^{\rho}$ & $G^a_{\mu\rho}$ & $0^{+(+)},\,1^{-(+)}$ \\
  $\gamma^{\rho}$ & $\tilde{G}^a_{\mu\rho}$ & $0^{-(+)},\,1^{+(+)}$ \\
  $\gamma^{\rho}\gamma_5$ & $G^a_{\mu\rho}$ & $0^{-(-)},\,1^{+(-)}$ \\
  $\gamma^{\rho}\gamma_5$ & $\tilde{G}^a_{\mu\rho}$ & $0^{+(-)},\,1^{-(-)}$ \\
\end{tabular}
\end{table}

\begin{figure}[htp!]
\subfloat[Diagram I (LO perturbation theory)]{%
    \includegraphics[width=.26\textwidth]{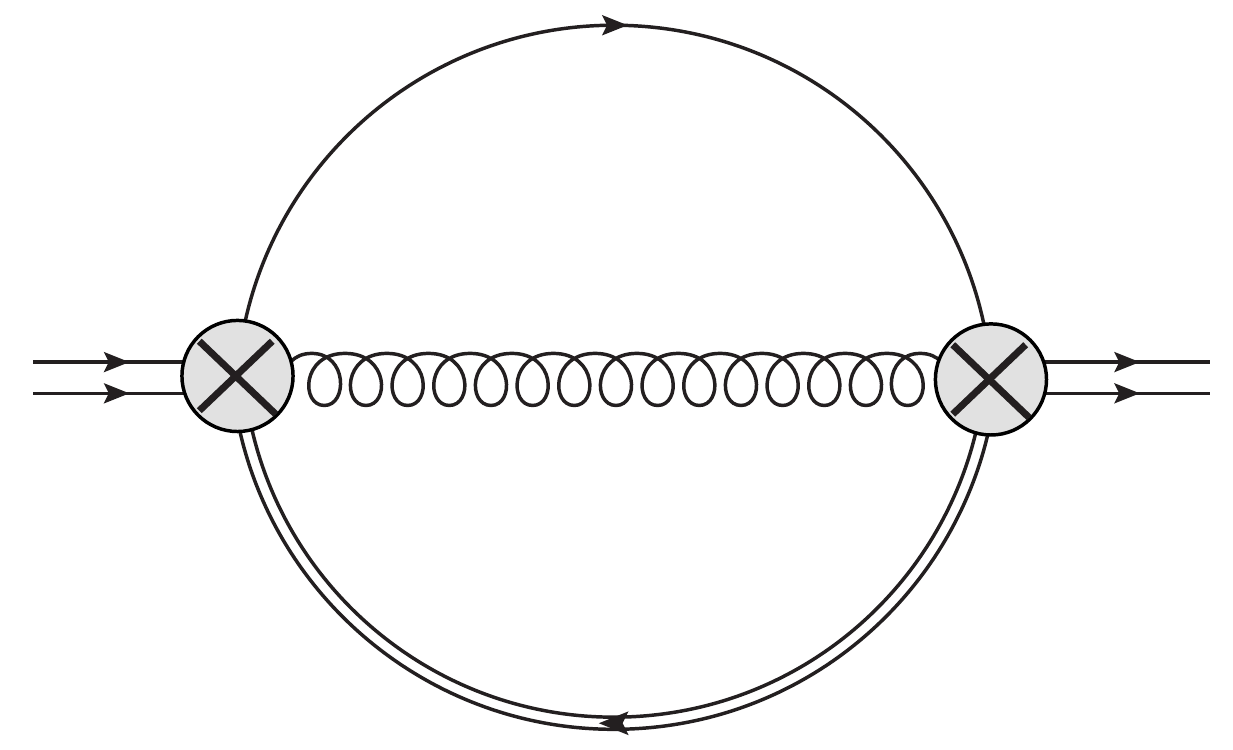}}\hfill
  \subfloat[Diagram II (dimension-four)]{%
    \includegraphics[width=.26\textwidth]{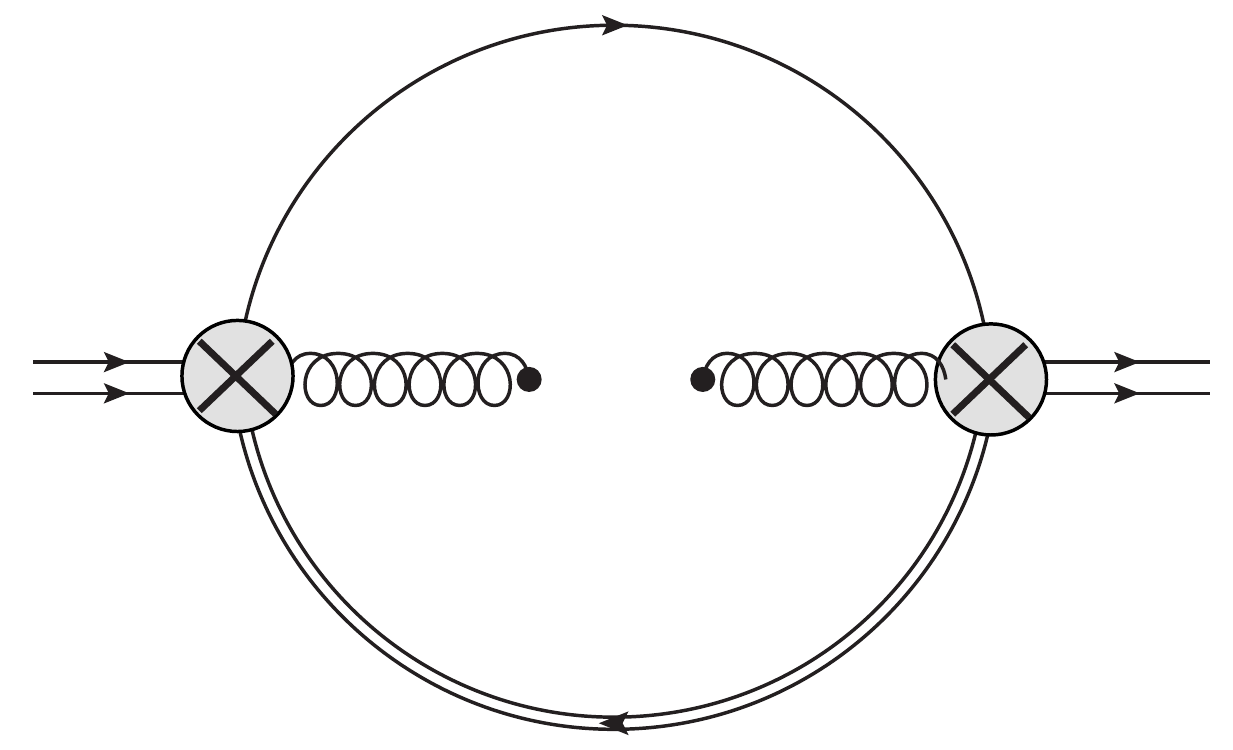}}\hfill
  \subfloat[Diagram III (dimension-four)]{%
    \includegraphics[width=.26\textwidth]{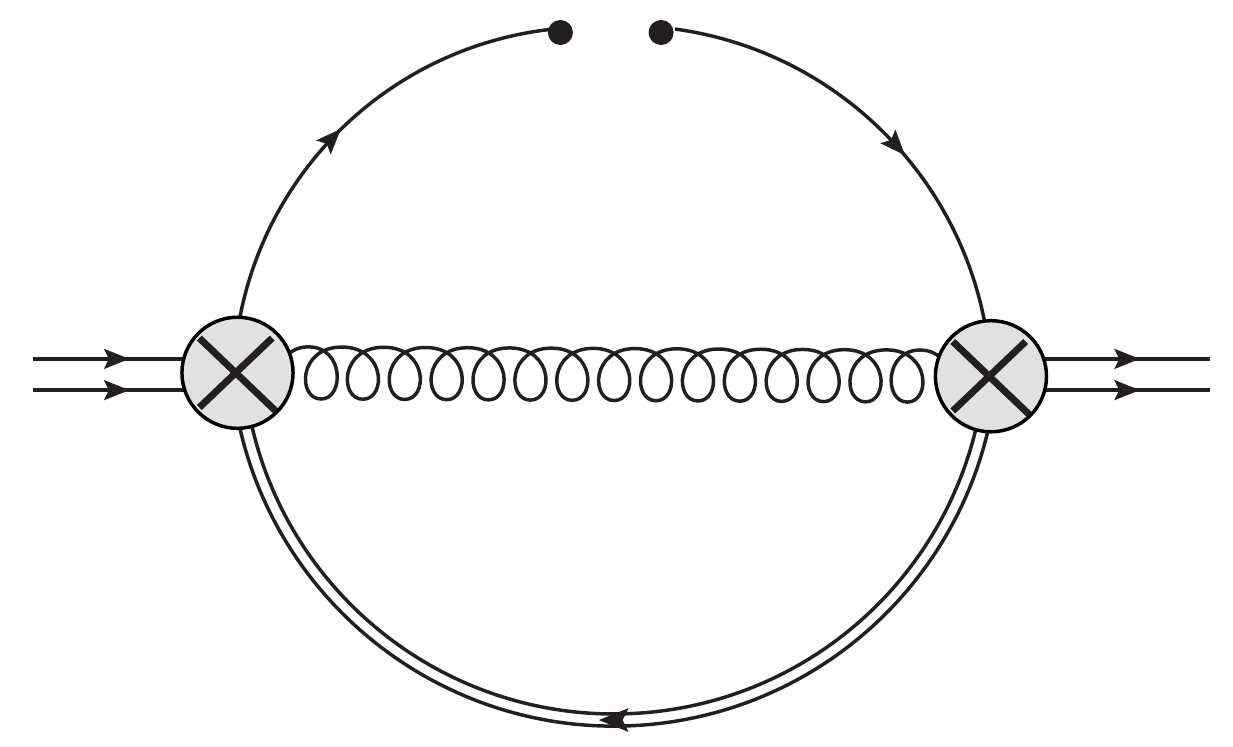}}\\
  \subfloat[Diagram V (dimension-six)]{%
  	\includegraphics[width=.26\textwidth]{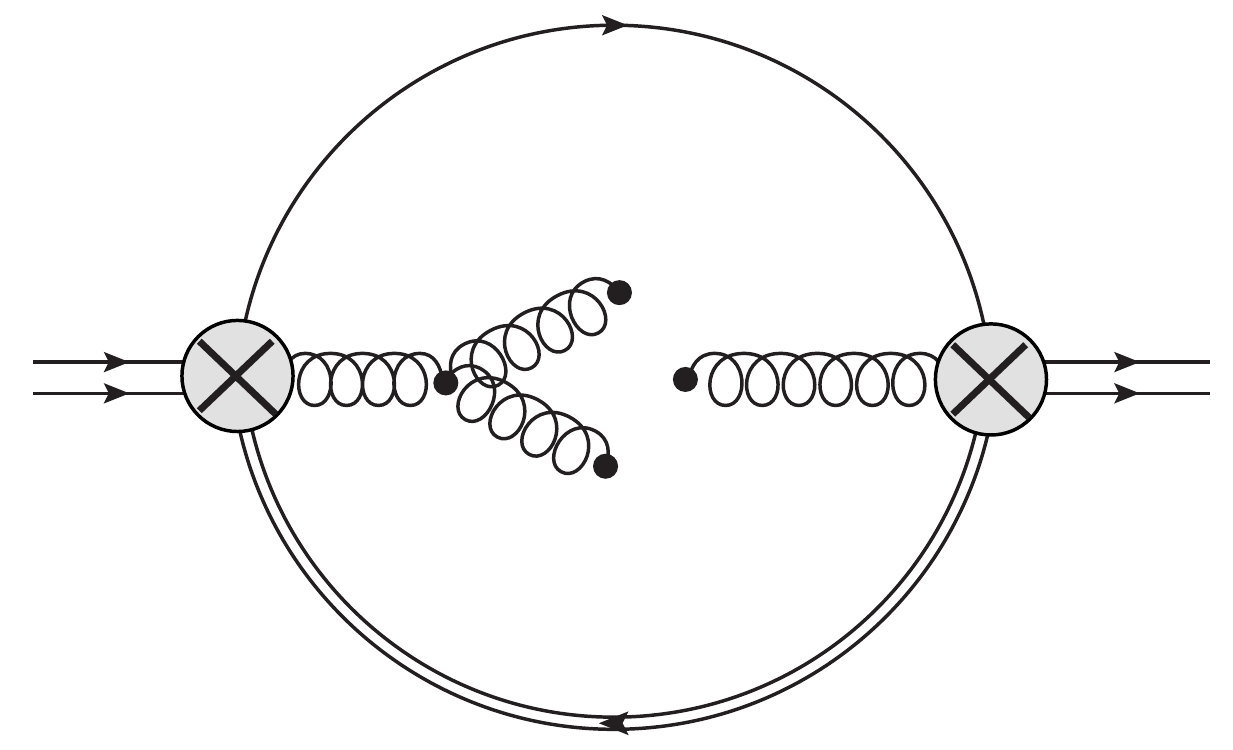}}\hfill
  \subfloat[Diagram VI (dimension-six)]{%
    \includegraphics[width=.26\textwidth]{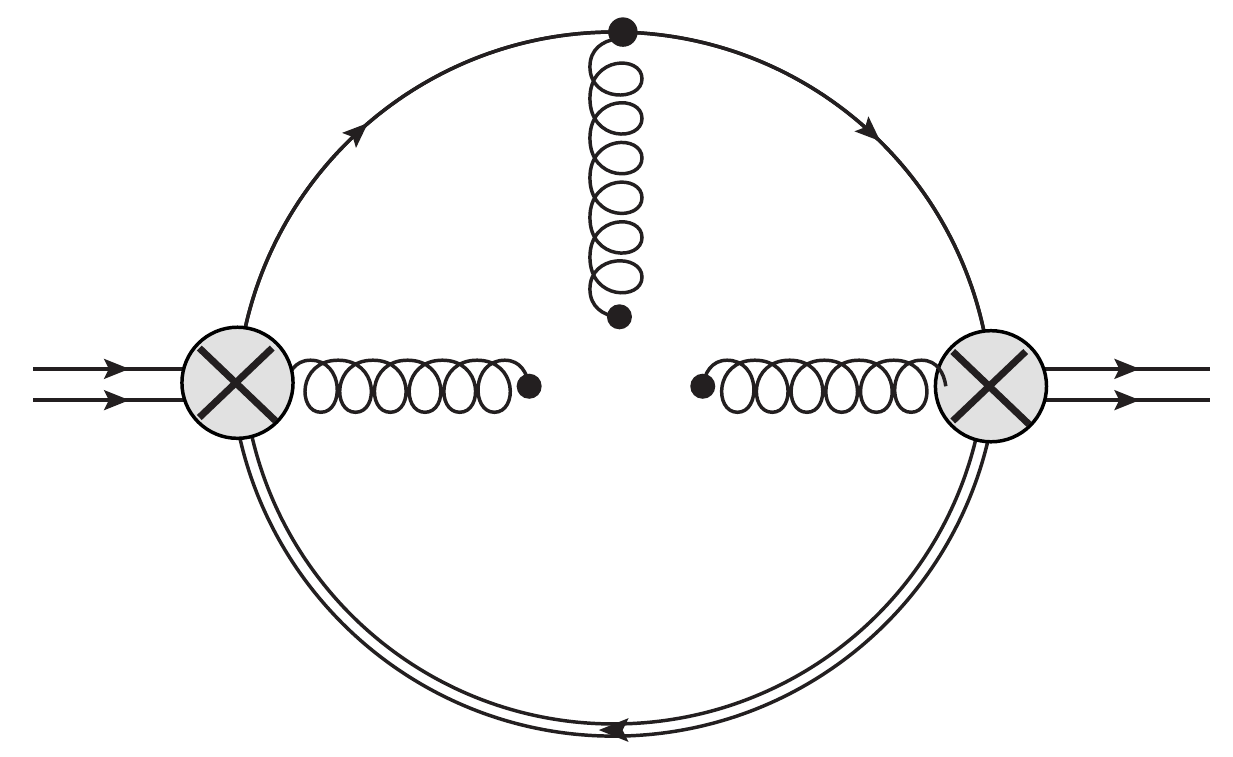}}\hfill
  \subfloat[Diagram VI (dimension-six)]{%
  	\includegraphics[width=.26\textwidth]{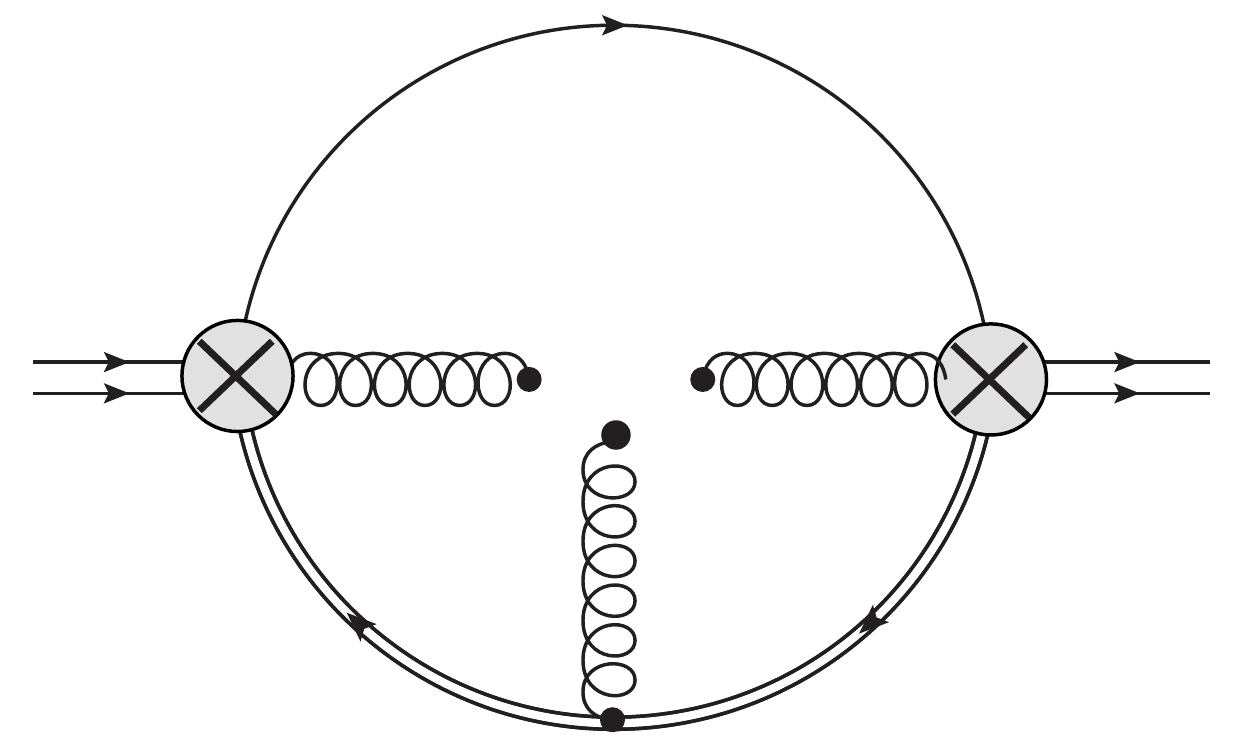}}\\
  \subfloat[Diagram VII (dimension-five)]{%
    \includegraphics[width=.26\textwidth]{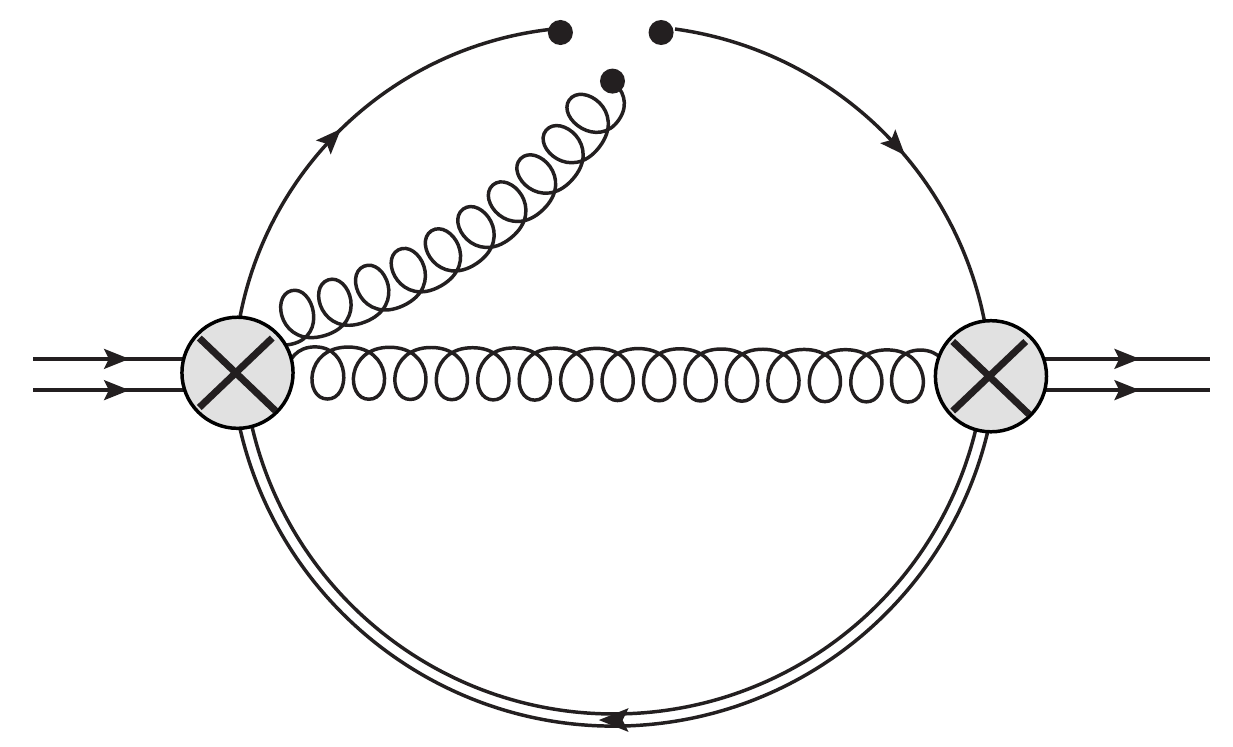}}\hfill
  \subfloat[Diagram VIII (dimension-five)]{%
    \includegraphics[width=.26\textwidth]{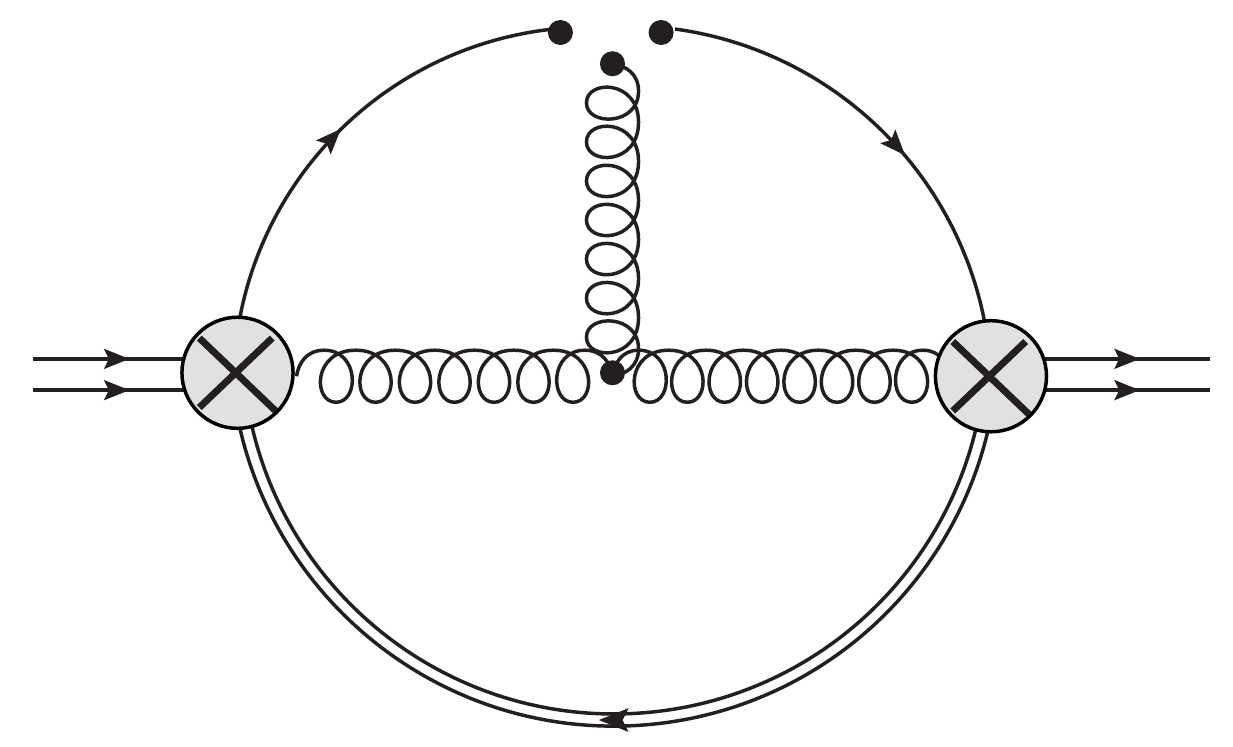}}\hfill
  \subfloat[Diagram IX (dimension-five)]{%
    \includegraphics[width=.26\textwidth]{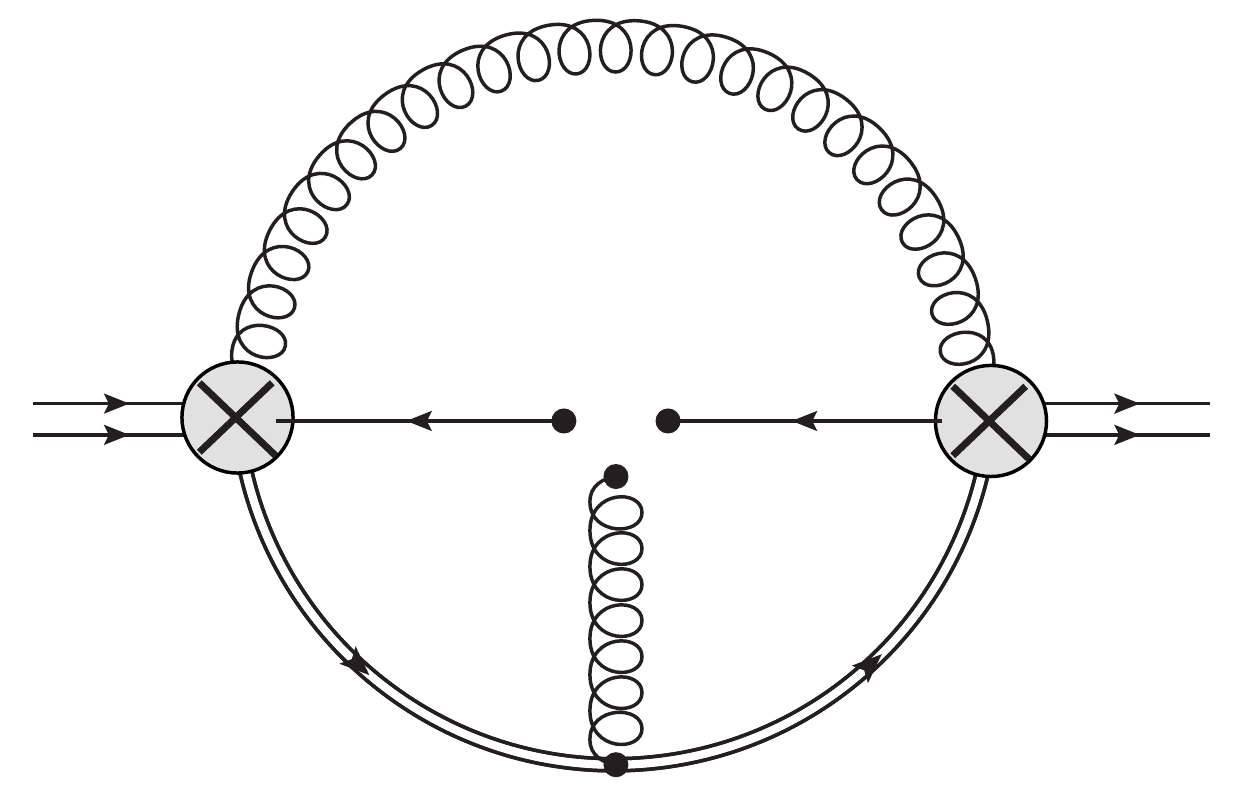}}\\
  \subfloat[Diagram X (dimension-five)]{%
    \includegraphics[width=.26\textwidth]{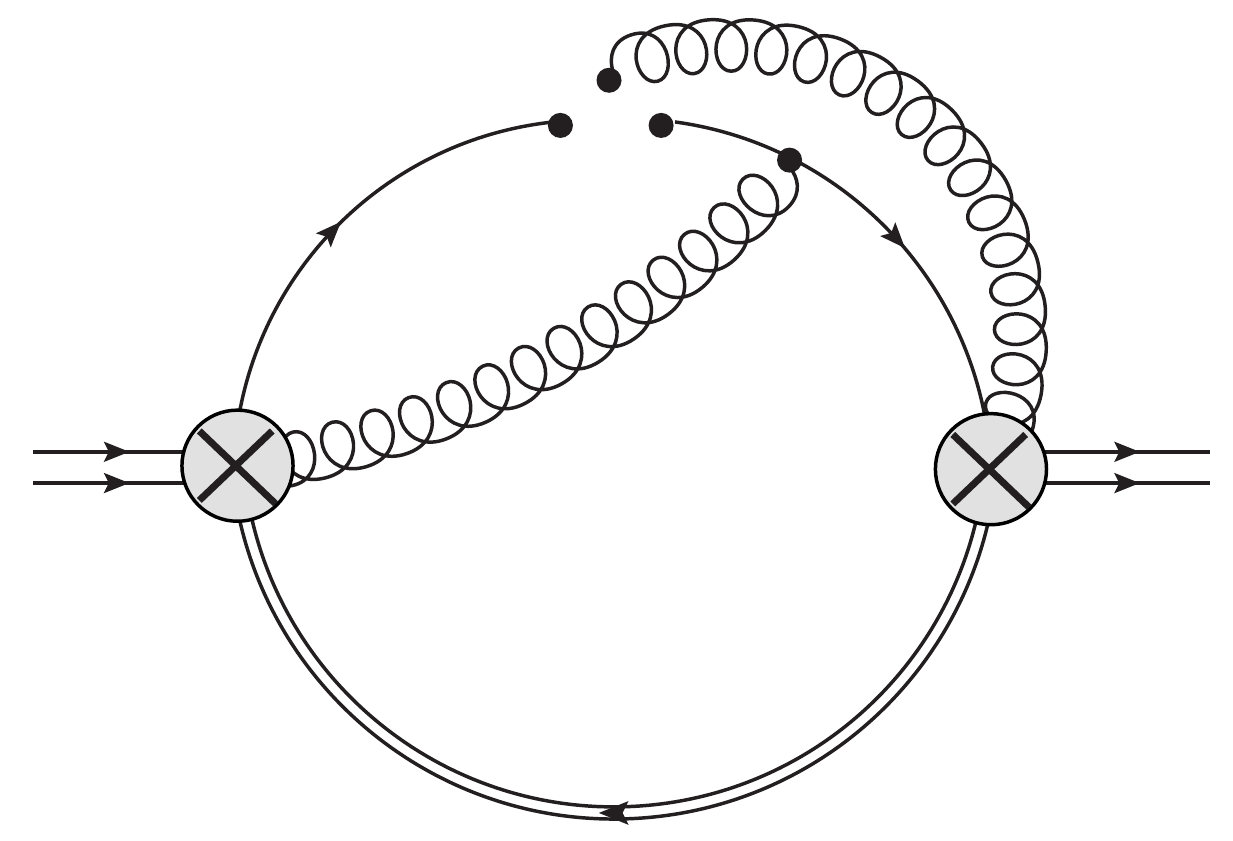}}\hfill
  \subfloat[Diagram XI (dimension-five)]{%
    \includegraphics[width=.26\textwidth]{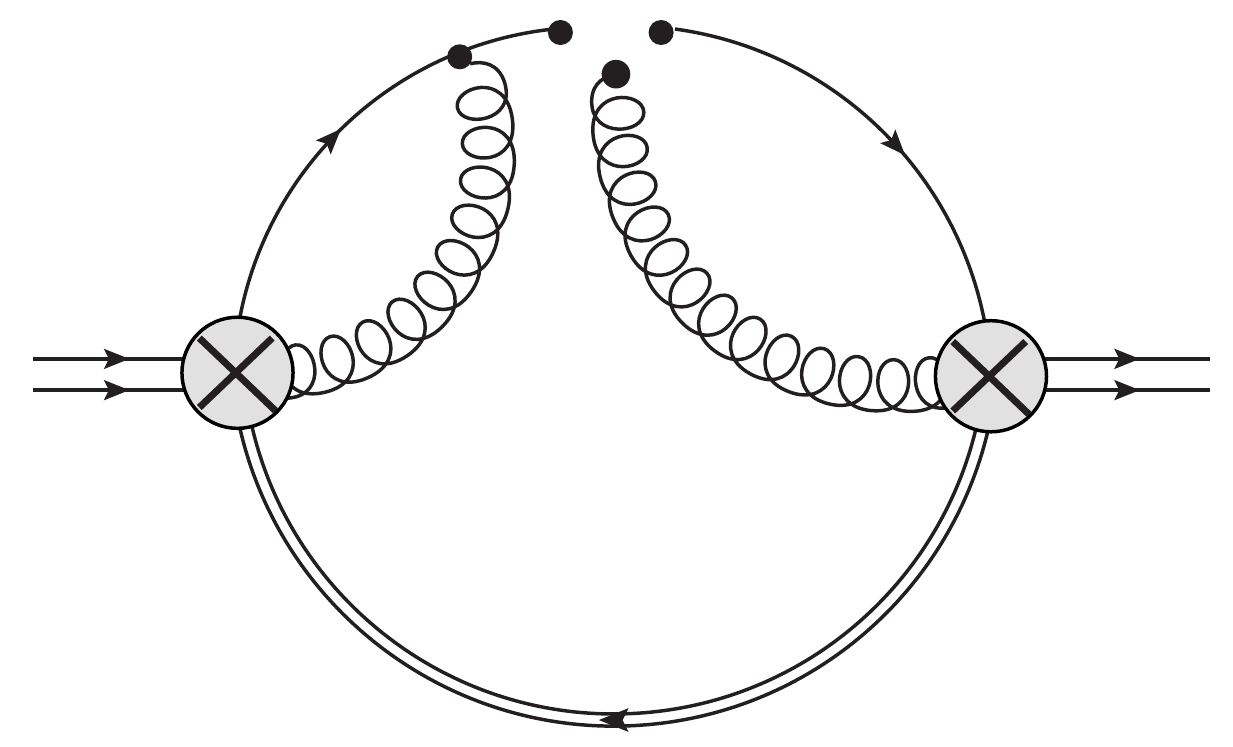}}\hfill
  \subfloat[Diagram XII (dimension-five)]{%
    \includegraphics[width=.28\textwidth]{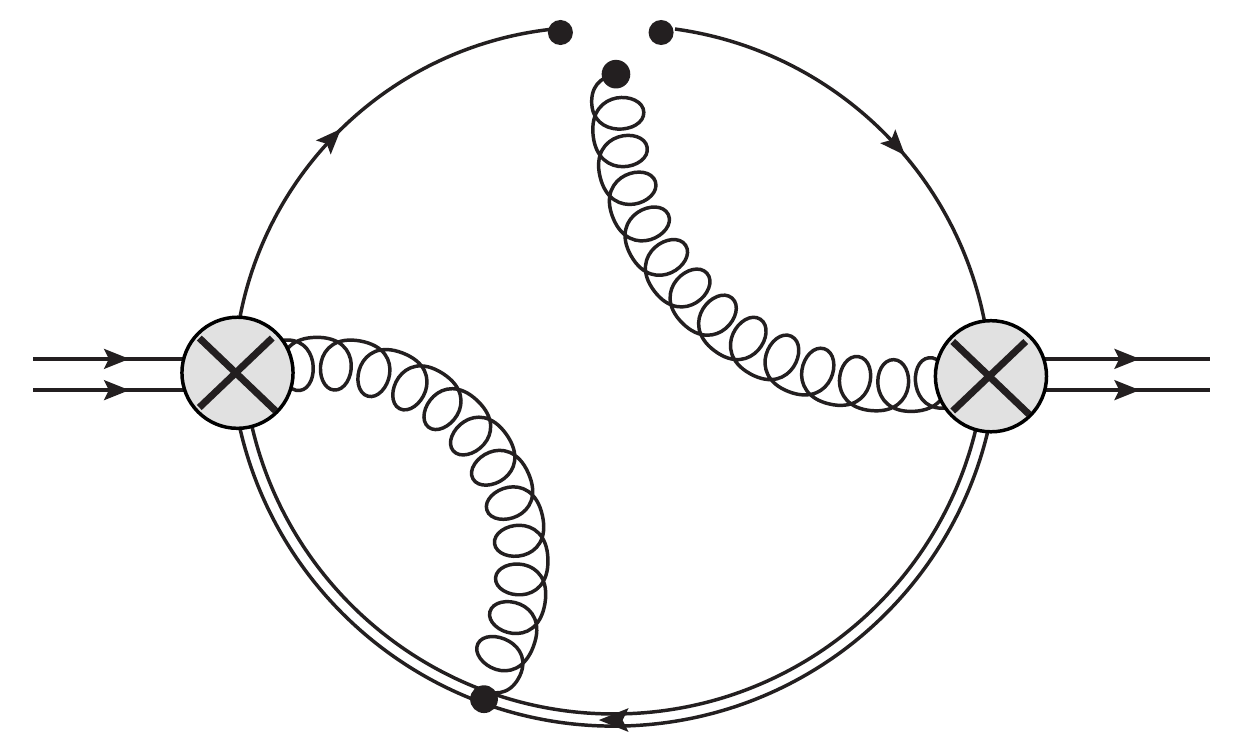}}\\
  \subfloat[Diagram XIII (dimension-five)]{%
    \includegraphics[width=.28\textwidth]{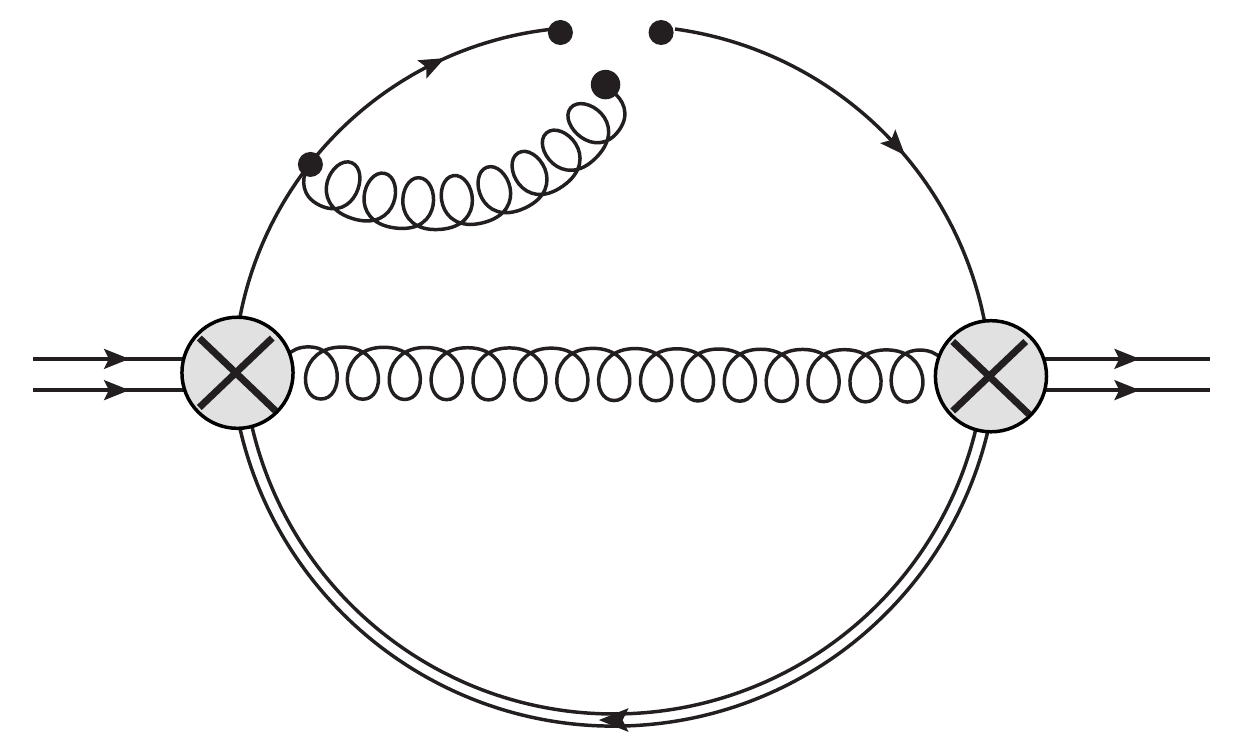}}\hfill
  \subfloat[Diagram XIV (dimension-five)]{%
    \includegraphics[width=.28\textwidth]{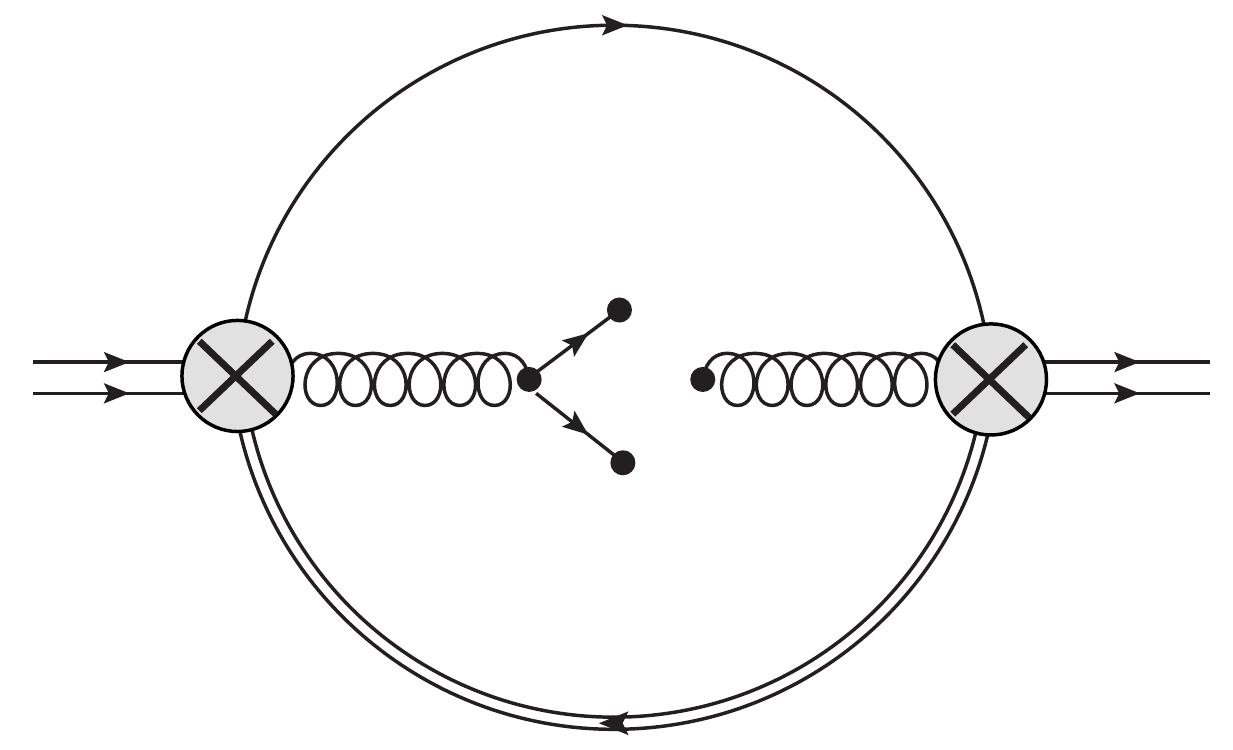}}\hfill
\caption{The Feynman diagrams calculated for the correlator~(\ref{correlator}).
  Single solid lines correspond to light quark propagators whereas
  double solid lines correspond to heavy quark propagators.\label{feynman_diagrams}}
\end{figure}

We calculate the correlators~(\ref{correlator}) within the operator product expansion (OPE)
in which perturbation theory is supplemented by a collection of non-perturbative terms,
each of which is the product of a perturbatively computed Wilson coefficient   
and a non-zero vacuum expectation value (VEV) corresponding to a QCD condensate.  
We include condensates up to 6d:
\begin{gather}
  \quarkthreed=\angled{\overline{q}_i^{\alpha} q_i^{\alpha}}
    \label{condensate_quark_three}\\
  \gluonfourd=\angled{\alpha_s G^a_{\mu\nu} G^a_{\mu\nu}}\label{condensate_gluon_four}\\
  \mixed=\angled{g_s \overline{q}_i^{\alpha}\sigma^{\mu\nu}_{ij}
    \lambda^a_{\alpha\beta} G^a_{\mu\nu} q_j^{\beta}}
    \label{condensate_mixed}\\
  \gluonsixd=\angled{g_s^3 f^{abc} G^a_{\mu\nu}G^b_{\nu\rho}G^c_{\rho\mu}}\,,\label{condensate_gluon_six}
\end{gather}
respectively referred to as
the 3d~quark condensate,
the 4d~gluon condensate,
the 5d~mixed condensate, and
the 6d~gluon condensate.
Superscripts on light quark fields are colour indices whereas subscripts are Dirac indices, and $\sigma^{\mu\nu}=\frac{i}{2}[\gamma^{\mu},\gamma^{\nu}]$.
The Wilson coefficients (including perturbation theory) are 
computed to leading-order (LO) in $g_s$
using coordinate-space fixed-point gauge techniques 
(see~\cite{BaganAhmadyEliasEtAl1994,PascualTarrach1984}, for example).
Note that LO contributions to~(\ref{correlator}) associated with 
6d~quark condensates are $\mathcal{O}(g_s^4)$; 
our calculation is actually $\mathcal{O}({g_s^3})$, 
and so 6d~quark condensates have been excluded 
from~(\ref{condensate_quark_three})--(\ref{condensate_gluon_six}).
(In Ref.~\cite{ChenKleivSteeleEtAl2013} the numerical effect of the 6d quark condensates has 
been shown to be small compared to the 6d gluon condensate).
Light quark mass effects are included in perturbation theory through a 
next-to-leading-order light quark mass expansion, 
and at leading-order in all other OPE terms.
The contributing Feynman diagrams are depicted in 
Figure~\ref{feynman_diagrams}\footnote{All Feynman diagrams are drawn using
JaxoDraw~\cite{BinosiTheussl2004}.}
where we follow as closely as possible the labeling scheme
of~\cite{LatorrePascualNarison1987}. 
(Note that there is no Diagram IV in 
Figure~\ref{feynman_diagrams} because, in~\cite{LatorrePascualNarison1987}, 
Diagram IV corresponds to an OPE contribution stemming from 
6d~quark condensates that is absent in the open-flavour heavy-light systems.)
The $\overline{\rm MS}$-scheme with the $D=4+2\epsilon$ convention is used, and $\mu$ 
is the corresponding renormalization scale.  
We use the program TARCER~\cite{MertigScharf1998},
which implements the recurrence algorithm of~\cite{Tarasov1996,Tarasov1997},
to express each diagram in terms of a small collection of 
master integrals, all of which are well-known.
Following~\cite{ChanowitzFurmanHinchliffe1979}, 
we employ a dimensionally regularized $\gamma_5$ that satisfies
$\gamma_5^2=1$ and $\{\gamma_5,\,\gamma^{\mu}\}=0$.
Note that the imaginary parts of Diagrams I--III were actually first computed 
between~\cite{GovaertsReindersRubinsteinEtAl1985} and GRW;
for these three diagrams, we were able to successfully bench-mark our results against
that original work.

\begin{figure}[ht!]
\centering
\subfloat[]{%
    \includegraphics[width=.26\textwidth]{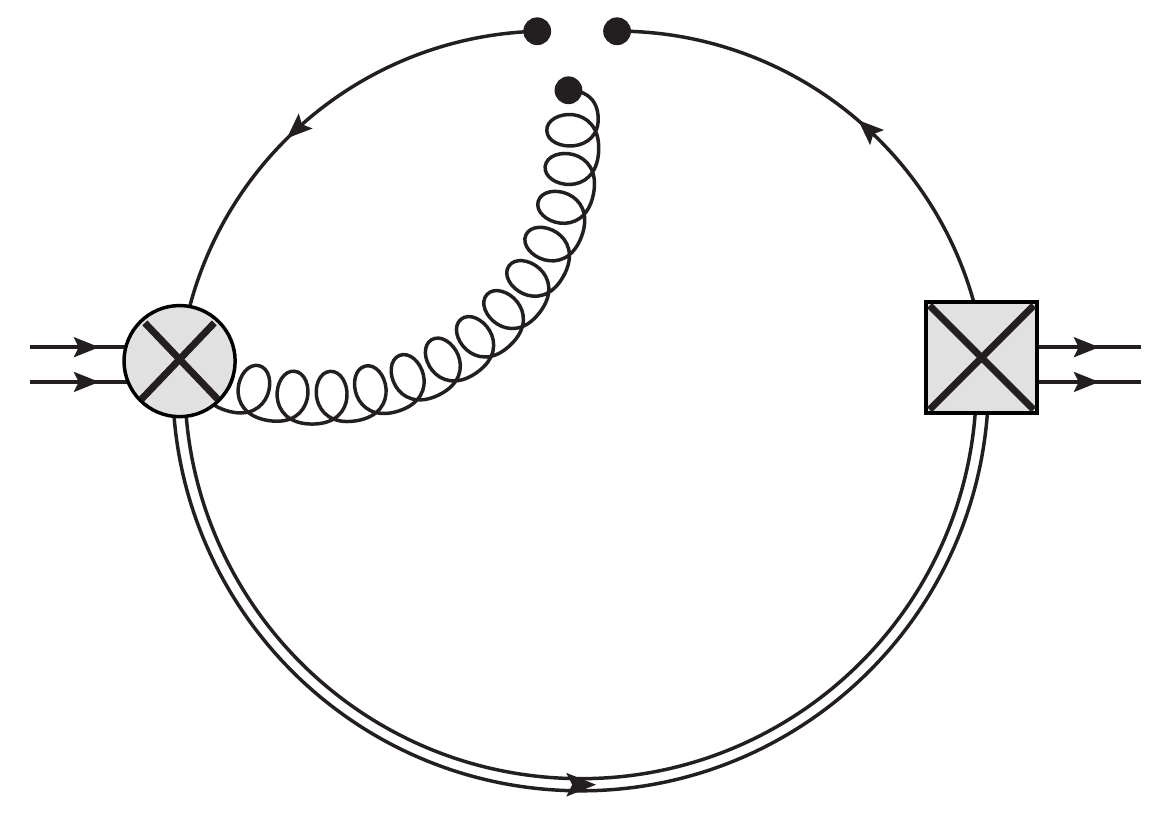}}
  \subfloat[]{%
    \includegraphics[width=.26\textwidth]{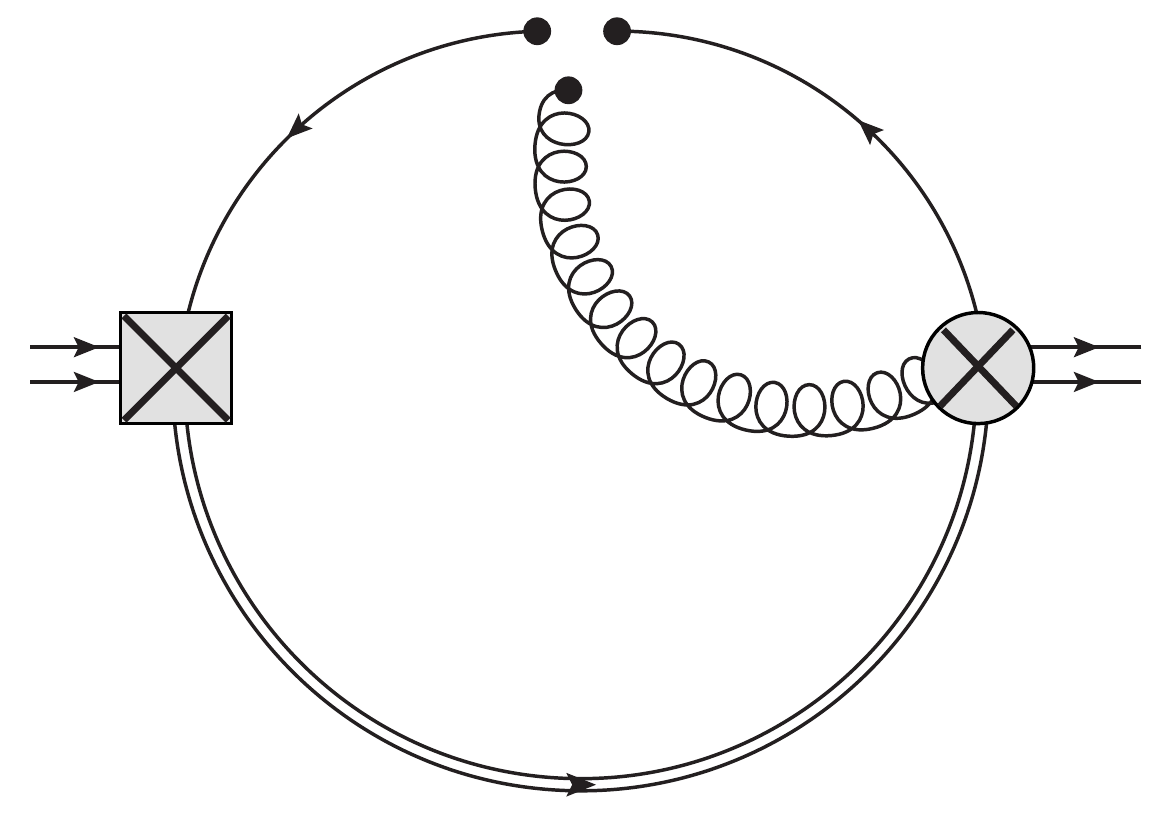}}
\caption{Renormalization-induced Feynman diagrams. Square insertion represents the mixing current.\label{renormalization_mixed}}
\end{figure}

Diagram~XII, a 5d mixed condensate contribution, generates some complications.  
Focusing on the lower portion of the diagram,  we see a heavy quark propagator carrying momentum $q$ 
multiplied by a divergent, one-loop sub-graph.  
Correspondingly, Diagram~XII contributes to the correlator a non-local divergence
proportional to
\begin{equation}\label{non_local}
  \frac{1}{(q^2-M_Q^2)\epsilon}.
\end{equation}
Following~\cite{ChenJinKleivSteeleWangXu2013},
this divergence is eliminated by renormalization of the composite operators~(\ref{current}) 
which induces mixing with either $\overline{Q}\gamma_{\mu}q$ 
or $\overline{Q}\gamma_{\mu}\gamma_5 q$.
Specifically, for those operators with $\Gamma^{\rho}=\gamma^{\rho}$ 
(recall~(\ref{Gamma_definition})), this mixing results in
\begin{equation}\label{vector}
  j_{\mu}\rightarrow j_{\mu} + k\frac{M_Q^2 \as}{\pi\epsilon}\overline{Q}\gamma_{\mu}q
\end{equation}
whereas, for those with $\Gamma^{\rho}=\gamma^{\rho}\gamma_5$, we have
\begin{equation}\label{axial}
  j_{\mu}\rightarrow j_{\mu} + k\frac{M_Q^2 \as}{\pi\epsilon}\overline{Q}\gamma_{\mu}\gamma_5 q
\end{equation}
where $k$ is an as yet undetermined constant emerging from renormalization.
For currents that mix according to~(\ref{vector}),
the VEV under the integral on the right-hand side of~(\ref{correlator}) 
gets modified as follows:
\begin{multline}
  \vev{\tau j_{\mu}(x)j^{\dag}_{\nu}(0)}\rightarrow
  \vev{\tau j_{\mu}(x)j^{\dag}_{\nu}(0)}\\
  +k\frac{M_Q^2 \as}{\pi\epsilon}\vev{\tau \overline{Q}(x)\gamma_{\mu}q(x) j_{\nu}^{\dag}(0)}
  +k^{*}\frac{M_Q^2 \as}{\pi\epsilon}\vev{\tau j_{\mu}(x)\overline{q}(0)\gamma_{\mu}Q(0)}
\label{correlator_renormalized}
\end{multline}
with an analogous expression for operators that mix according to~(\ref{axial}).
The first term on the right-hand side of~(\ref{correlator_renormalized}) corresponds
to the diagrams of Figure~\ref{feynman_diagrams} whereas the last two terms give
rise to new, renormalization-induced contributions to the OPE.
Almost all of these new contributions are sub-leading in $g_s$, 
however, and so are ignored.  The only exceptions are those containing 
the 5d~mixed condensate~(\ref{condensate_mixed}); these give rise to the pair
of diagrams depicted in Figure~\ref{renormalization_mixed}.  
Both of these tree-level diagrams contain a heavy quark propagator with
momentum $q$ and are multiplied
by a factor of $\frac{1}{\epsilon}$ in~(\ref{correlator_renormalized}),
precisely what is needed to cancel the non-local divergence~(\ref{non_local})

Summing the diagrams from Figures~\ref{feynman_diagrams} and~\ref{renormalization_mixed},
and then determining the constant $k$ from~(\ref{vector}) or~(\ref{axial}) such that all non-local
divergences are eliminated, we find for either $\Pi^{(0)}$ or $\Pi^{(1)}$ 
from~(\ref{spin_breakdown}) that
\begin{multline}
  \Pi(q^2)=
  \frac{M_Q^6 \as}{960\pi^3 z^2}\left[f^{(\text{pert})}_1(z)\log(1-z) 
    + f^{(\text{pert})}_2(z)\dilog(z) + c^{(\text{pert})}z\right]\\
    +\frac{M_Q^5 m_q\as}{\pi^3 z^2}\left[f^{(m)}_1(z)\log(1-z) 
    + f^{(m)}_2(z)\dilog(z) + c^{(m)}z\right]\\
  +\frac{M_Q^3 \as\quarkthreed}{6\pi z^2}\left[f^{(qq)}(z)\log(1-z) + c^{(qq)}z\right]\\
  +\frac{M_Q^2\gluonfourd}{144\pi z^2}\left[f^{(GG)}(z)\log(1-z) + c^{(GG)}z\right]\\
  +\frac{M_Q\as\mixed}{3456\pi z^2}\left[f^{(qGq)}_1(z)\log(1-z) 
    + f^{(qGq)}_2(z)\frac{z^2}{1-z}\log\left(\frac{M_Q^2}{\mu^2}\right)
    + f^{(qGq)}_3(z)\frac{z}{(1-z)}\right]\\
  +\frac{\gluonsixd}{192\pi^2 z^2}\left[f^{(GGG)}(z)\log(1-z) + c^{(GGG)}z\right]\\
  \label{correlator_result}
\end{multline}
where
\begin{equation}\label{z_definition}
  z=\frac{q^2}{M_Q^2}
\end{equation}
and where $\dilog(z)$ is the dilogarithm function defined by
\begin{equation}
\dilog(z)=\int_{z}^{\infty}\frac{\mathrm{ln}(1-t)}{t}\dt.
\end{equation}
The remaining quantities in~\eqref{correlator_result}
are listed in Tables~\ref{pert_table}--\ref{glue_sixd_table} 
for the distinct $J^{P(C)}$ combinations under consideration.
Also, in Table~\ref{a_table}, we give the values determined for the renormalization 
parameter $k$.
Finally, we note that, for the sake of brevity, we have omitted all polynomials in $z$ corresponding to dispersion-relation subtractions
from~(\ref{correlator_result}) and Tables \ref{pert_table} to \ref{glue_sixd_table}. As discussed in Section \ref{III}, these subtraction constants do not contribute to the Laplace sum-rules. 
\begin{table}
\centering
\caption{The polynomials and constants of perturbation theory.}
\label{pert_table}
\begin{tabular}{cccc}
  $J$ & $f^{(\text{pert})}_1(z)$ & $f^{(\text{pert})}_2(z)$ & $c^{(\text{pert})}$\\
\hline
  $0$ & $-3+30z+20z^2-60z^3+15z^4-2z^5$ & $-60z^2$ & -3\\
  $1$ & $-1-140z^2+100z^3+45z^4-4z^5$ & $60z^2 (1+2z)$& -1\\
\end{tabular}
\end{table}

\begin{table}
\centering
\caption{The polynomials and constants of the light quark mass correction to perturbation theory.}
\label{pert_mass_table}
\begin{tabular}{cccc}
  $J^{P(C)}$ & $f^{(m)}_1(z)$ & $f^{(m)}_2(z)$ & $c^{(\text{m})}$\\
\hline
  $0^{+(+)}$ & $3(1-6z+18z^2-10z^3-3z^4)$ & $-36z^3$ & 3\\
  $0^{-(-)}$ & $-3(1-6z+18z^2-10z^3-3z^4)$ & $36z^3$ & -3\\
  $0^{-(+)}$ & $3(1-6z+18z^2-10z^3-3z^4)$ & $-36z^3$ & 3\\
  $0^{+(-)}$ & $-3(1-6z+18z^2-10z^3-3z^4)$ & $36z^3$ & -3\\
  $1^{+(+)}$ & $1-6z+18z^2-10z^3-3z^4$ & $-12z^3$ & 1\\
  $1^{-(-)}$ & $-(1-6z+18z^2-10z^3-3z^4)$ & $12z^3$ & -1\\
  $1^{-(+)}$ & $1-6z+18z^2-10z^3-3z^4$ & $-12z^3$ & 1\\
  $1^{+(-)}$ & $-(1-6z+18z^2-10z^3-3z^4)$ & $12z^3$ & -1\\
\end{tabular}
\end{table}

\begin{table}
\centering\caption{The polynomials and constants of the 3d quark condensate contribution.}
\label{quark_threed_table}
\begin{tabular}{ccc}
  $J^{P(C)}$ & $f^{(qq)}(z)$ & $c^{(qq)}$\\
\hline
  $0^{+(+)}$ & $-(1-z)^3$ & -1\\
  $0^{-(-)}$ & $(1-z)^3$ & 1\\
  $0^{-(+)}$ & $-(1-z)^3$ & -1\\
  $0^{+(-)}$ & $(1-z)^3$ & 1\\
  $1^{+(+)}$ & $-\frac{1}{3}(1-z)^3$ & $-\frac{1}{3}$\\
  $1^{-(-)}$ & $\frac{1}{3}(1-z)^3$ & $\frac{1}{3}$\\
  $1^{-(+)}$ & $-\frac{1}{3}(1-z)^3$ & $-\frac{1}{3}$\\
  $1^{+(-)}$ & $\frac{1}{3}(1-z)^3$ & $\frac{1}{3}$\\
\end{tabular}
\end{table}

\begin{table}
\centering
\caption{The polynomials and constants of the 4d gluon condensate contribution.}
\label{glue_fourd_table}
\begin{tabular}{ccc}
  $J^{P(C)}$ & $f^{(GG)}(z)$ & $c^{(GG)}$\\
\hline
  $0^{+(+)}$ & $3(1-z)^2 (1+2z)$ & 3\\
  $0^{-(-)}$ & $3(1-z)^2 (1+2z)$ & 3\\
  $0^{-(+)}$ & $-3(1-z)^2 (1+2z)$ & -3\\
  $0^{+(-)}$ & $-3(1-z)^2 (1+2z)$ & -3\\
  $1^{+(+)}$ & $-(1-z)^2 (1-4z)$ & -1\\
  $1^{-(-)}$ & $-(1-z)^2 (1-4z)$ & -1\\
  $1^{-(+)}$ & $(1-z)^2 (1-4z)$ & 1\\
  $1^{+(-)}$ & $(1-z)^2 (1-4z)$ & 1\\
\end{tabular}
\end{table}

\begin{table}
\centering
\caption{The polynomials and constants of the 5d mixed condensate contribution.}
\label{mixed_fived_table}
\begin{tabular}{cccc}
  $J^{P(C)}$ & $f^{(qGq)}_1(z)$ & $f^{(qGq)}_2(z)$ & $f^{(qGq)}_3(z)$\\
\hline
  $0^{+(+)}$ & $18(1-z)(13+2z)$  & $-36(17-z)$   & $9(26+27z-21z^2)$ \\
  $0^{-(-)}$ & $-18(1-z)(13+2z)$ & $36(17-z)$    & $-9(26+27z-21z^2)$ \\
  $0^{-(+)}$ & $-18(1-z)(27+2z)$ & $-36(7+z)$    & $-3(162-351z+29z^2)$ \\
  $0^{+(-)}$ & $18(1-z)(27+2z)$  & $36(7+z)$     & $3(162-351z+29z^2)$ \\
  $1^{+(+)}$ & $-6(1-z)(27-38z)$ & $12(21+19z)$  & $-(162+369z+205z^2)$ \\
  $1^{-(-)}$ & $6(1-z)(27-38z)$  & $-12(21+19z)$ & $162+369z+205z^2$ \\
  $1^{-(+)}$ & $6(1-z)(13-38z)$  & $12(51-19z)$  & $78-999z+569z^2$ \\
  $1^{+(-)}$ & $-6(1-z)(13-38z)$ & $-12(51-19z)$ & $-(78-999z+569z^2)$ \\
\end{tabular}
\end{table}

\begin{table}
\centering
\caption{The polynomials and constants of the 6d gluon condensate contribution.}
\label{glue_sixd_table}
\begin{tabular}{ccc}
  $J^{P(C)}$ & $f^{(GGG)}(z)$ & $c^{(GGG)}$\\
\hline
  $0^{+(+)}$ & -3 & -3 \\
  $0^{-(-)}$ & -3 & -3 \\
  $0^{-(+)}$ & 3 & 3 \\
  $0^{+(-)}$ & 3 & 3 \\
  $1^{+(+)}$ & $1-2z$ & 1 \\
  $1^{-(-)}$ & $1-2z$ & 1 \\
  $1^{-(+)}$ & $-(1-2z)$ & -1 \\
  $1^{+(-)}$ & $-(1-2z)$ & -1 \\
\end{tabular}
\end{table}

\begin{table}
\centering
\caption{The renormalization parameter $k$ from~(\ref{vector}) and~(\ref{axial}).}
\label{a_table}
\begin{tabular}{cc}
  $J^{P(C)}$ & $k$ \\
\hline
  $0^{+(+)}$ & $-\frac{2i}{3}$ \\
  $0^{-(-)}$ & $-\frac{2i}{3}$ \\
  $0^{-(+)}$ & $-\frac{1}{3}$ \\
  $0^{+(-)}$ & $-\frac{1}{3}$ \\
  $1^{+(+)}$ & $-\frac{5}{9}$ \\
  $1^{-(-)}$ & $-\frac{5}{9}$ \\
  $1^{-(+)}$ & $-\frac{4i}{9}$ \\
  $1^{+(-)}$ & $-\frac{4i}{9}$ \\
\end{tabular}
\end{table}

\section{QCD Laplace Sum-Rules}\label{III}
Viewed as a function of Euclidean momentum $Q^2=-q^2$, 
each of $\Pi^{(0)}$ and $\Pi^{(1)}$ from~(\ref{spin_breakdown})
satisfies a dispersion relation
\begin{equation}\label{dispersion_relation}
  \Pi(Q^2)=\frac{Q^8}{\pi}\int_{t_0}^{\infty}
  \frac{\Im\Pi(t)}{t^4(t+Q^2)}
  \,\dt+\cdots,\ Q^2>0
\end{equation}
where $\cdots$ represents subtractions constants, collectively a third degree polynomial in $Q^2$, 
and $t_0$ represents the appropriate physical threshold.
The quantity $\Pi$ on the left-hand side of~(\ref{dispersion_relation}) is identified with 
the OPE result~(\ref{correlator_result}) while
$\Im\Pi$ on the right-hand side of~(\ref{dispersion_relation}) is the hadronic spectral function.
To eliminate the (generally unknown) subtraction constants and enhance the ground state contribution
to the integral, the Borel transform 
\begin{equation}\label{borel}
  \borel =\lim_{\stackrel{N,Q^2\rightarrow\infty}{\tau=N/Q^2}}
  \frac{(-Q^2)^N}{\Gamma(N)}\left(\frac{d}{dQ^2}\right)^N
\end{equation}
is applied to~(\ref{dispersion_relation}) weighted by $(-Q^2)^k$ for $k\geq0$
to yield the $k^{\text{th}}$-order Laplace sum-rule (LSR)~\cite{ShifmanVainshteinZakharov1979}
\begin{equation}
\label{lsr_main}
  \lsr_k(\tau) =\int_{t_0}^{\infty}t^k e^{-t\tau}\frac{1}{\pi}\Im\Pi(t)\,\dt, 
  ~\lsr_k(\tau) =\frac{1}{\tau}\borel\left\{(-Q^2)^k\Pi(Q^2)\right\}.
\end{equation}
The Borel transform annihilates polynomials in $Q^2$ which eliminates dispersion-relation 
subtraction constants and  justifies our omission of 
polynomials (divergent or not) from~(\ref{correlator_result}).
The exponential kernel on the  right-hand side of~(\ref{lsr_main})
suppresses contributions from excited resonances and the continuum relative to the ground state.

In a typical QCD sum-rules analysis,
the hadronic spectral function is parametrized using a small number 
of hadronic quantities, predictions for which are then extracted using
a fitting procedure.
We employ the ``single narrow resonance plus continuum'' model
\cite{ShifmanVainshteinZakharov1979}
\begin{equation}\label{single_narrow_resonance}
  \frac{1}{\pi}\Im\Pi(t) = f_H^2 m_{H}^{8} \delta(t-m_H^2)
  +\theta(t-s_0)\frac{1}{\pi}\Im\Pi^{\text{OPE}}(t)
\end{equation}
where $m_H$ is the ground state resonance mass, $f_H$ is its coupling strength,
$\theta$ is a Heaviside step function, $s_0$ is the continuum threshold
and $\Im\Pi^{\text{OPE}}$ is the imaginary part of 
the QCD expression for $\Pi$ given in~(\ref{correlator_result}).
Substituting~(\ref{single_narrow_resonance}) into~(\ref{lsr_main}) gives
\begin{equation}\label{lsr_minus_continuum}
  \lsr_k(\tau) = f_H^2 m_H^{8+2k} e^{-m_H^2 \tau}
  + \int_{s_0}^{\infty}t^k e^{-t\tau}\frac{1}{\pi}\Im\Pi^{\text{OPE}}(t)\dt,
\end{equation}
and, defining continuum-subtracted LSRs by
\begin{equation}\label{lsr_subtracted}
  \lsr_k(\tau,\,s_0)=\lsr_k(\tau)
  -\int_{s_0}^{\infty}t^k e^{-t\tau}\frac{1}{\pi}\Im\Pi^{\text{OPE}}(t)\dt,
\end{equation}
we find, between~(\ref{lsr_minus_continuum}) and~(\ref{lsr_subtracted}), the result
\begin{equation}\label{lsr_main_subtracted}
  \lsr_k(\tau,\,s_0 ) = f_H^2 m_H^{8+2k} e^{-m_H^2 \tau}.
\end{equation}
Finally, using~(\ref{lsr_main_subtracted}), we obtain
\begin{equation}\label{lsr_master}
  \frac{\lsr_1(\tau,\,s_0 )}{\lsr_0(\tau,\,s_0 )}=m_H^2,
\end{equation}
the central equation of our analysis methodology.

To develop an OPE expression for $\lsr_k(\tau,\,s_0)$, we exploit a relationship
between the Borel transform and the inverse Laplace 
transform $\hat{\mathcal{L}}^{-1}$~\cite{ShifmanVainshteinZakharov1979}
\begin{equation}\label{borel_laplace_identity}\begin{split}
\frac{1}{\tau}\borel\left\{f(Q^2)\right\} &= \hat{\mathcal{L}}^{-1}\left\{f(Q^2)\right\}\\
               &= \frac{1}{2\pi i}\int_{c-i\infty}^{c+i\infty}f(Q^2) e^{Q^2 \tau} \mathrm{d}Q^2
\end{split}\end{equation}
where $c$ is chosen such that $f(Q^2)$ is analytic to the right of 
the integration contour in the complex $Q^2$-plane.
Applying definitions~(\ref{lsr_main}) and~(\ref{lsr_subtracted}) to~(\ref{correlator_result})
and using~(\ref{borel_laplace_identity}), it is straightforward  to show that
\begin{multline}\label{lsr_zero_ope}
  \lsr_0(\tau,\,s_0) = M_Q^2\int_1^{\frac{s_0}{M_Q^2}} e^{-x M_Q^2 \tau}
    \frac{1}{\pi}\Im\Pi^{\text{OPE}}(x M_Q^2)\,\dx\\
    +\frac{e^{-M_Q^2 \tau}M_Q^3\as\mixed}{108\pi}\left[a\log\left(\frac{M_Q^2}{\mu^2}\right)+b\right]
\end{multline}
and
\begin{equation}
  \lsr_1(\tau,\,s_0) = -\frac{d}{d\tau}\lsr_0(\tau,\,s_0) 
\end{equation}
where $a,\,b$ are constants given in Table~\ref{ab_table} for each $J^{P(C)}$ 
combination under investigation.
Note that the definite integral in~(\ref{lsr_zero_ope}) can be evaluated exactly;
however, the result is rather long and not particularly illuminating,
and so is omitted for brevity.

\begin{table}
\centering
\caption{The constants $a$ and $b$ from~(\ref{lsr_zero_ope}).}
\label{ab_table}
\begin{tabular}{ccc}
  $J^{P(C)}$ & a & b \\ 
\hline
  $0^{+(+)}$ & -18 & 9 \\
  $0^{-(-)}$ & 18 & -9 \\
  $0^{-(+)}$ & 9 & -15 \\
  $0^{+(-)}$ & -9 & 15 \\
  $1^{+(+)}$ & 15 & -23 \\
  $1^{-(-)}$ & -15 & 23 \\
  $1^{-(+)}$ & 12 & -11 \\
  $1^{+(-)}$ & -12 & 11 \\
\end{tabular}
\end{table}

Renormalization-group (RG) improvement~\cite{NarisondeRafael1981} 
dictates that the coupling constant and quark masses
in~(\ref{lsr_zero_ope}) be replaced by their (one-loop, \msbar) running counterparts.
The running coupling is given by
\begin{equation}\label{running_coupling}
  \as(\mu)=\frac{\as(M_X)}%
    {1+\frac{1}{12\pi}\left(33-2N_F\right)\as(M_X)\log\left(\frac{\mu^2}{M_X^2}\right)}
\end{equation}
where $N_F$ is the number of active quark flavors and
$M_X$ is a reference scale for experimental values of $\as$.
In addition,
the running heavy quark mass can be expressed as
\begin{equation}\label{running_heavy_quark}
  M(\mu)=M(\overline{M})\left[\frac{\as(\mu)}{\as(\overline{M})}\right]^{\frac{12}{33-2N_f}}
\end{equation}
where $\overline{M}$ is defined by $M(\overline{M})=\overline{M}$,
and the running light quark mass can be expressed as
\begin{equation}\label{running_light_quark}
  m(\mu)=m(2~\gev)\left[\frac{\as(\mu)}{\as(2~\gev)}\right]^{\frac{12}{33-2N_f}},
\end{equation}
in anticipation of using the Ref.~\cite{OliveEtAl2014} light-quark 
mass values at 2 GeV.
For charm systems, we use the renormalization scale $\mu=\overline{M}=M_c$
while for bottom systems $\mu=\overline{M}=M_b$ 
with PDG values~\cite{OliveEtAl2014}  
\begin{equation}
 M_c=(1.275 \pm 0.025)\ \gev\,,~M_b=(4.18 \pm 0.03)\ \gev.
\end{equation}
We then evaluate $\alpha_s$ via \eqref{running_coupling}
within the relevant flavour thresholds 
using appropriate Ref.~\cite{OliveEtAl2014}  
reference values at the $\tau$ and $Z$ masses
\begin{equation}
\as(M_{\tau})=0.330\pm0.014\,,~\as(M_Z)=0.1185\pm0.0006.
\end{equation}
Lastly, we use the following values for the light quark 
masses~\cite{OliveEtAl2014} 
\begin{gather}
  m_n(2~\gev)=\frac{1}{2}
  \left[m_u(2~\gev)+m_d(2~\gev) \right]=(3.40 \pm 0.25)\ \mev\,,\\
  m_s(2~\gev)=(93.5 \pm 2.5)\ \mev\,.
\end{gather}
The QCD predictions \eqref{correlator_result} have isospin symmetry because 
$\langle \bar u u\rangle=\langle\bar d d\rangle=\langle\bar n n\rangle$ and the 
sub-leading effect of nonstrange quark masses is negligible (i.e., we are 
effectively in the chiral limit for nonstrange systems). 

In addition to specifying expressions for the running coupling and quark masses,
we must also specify the numerical values of the 
condensates~(\ref{condensate_quark_three})--(\ref{condensate_gluon_six}).
Because of the form of~\eqref{correlator_result}, for 
$\quarkthreed$ we consider the product
\begin{equation}
  M\quarkthreed = \left(\frac{M}{m}\right)\angled{m \overline{q}q}
  \label{RGinvarQQ}
\end{equation}
as both $\frac{M}{m}$ and $\angled{m \overline{q}q}$ are RG-invariant  
quantities.
From PCAC~\cite{GMOR1968}  (using Ref.~\cite{Narison2004} conventions), we have
\begin{gather}
  \angled{m_{n}\overline{n}n} = -\frac{1}{2} f_{\pi}^2 m_{\pi}^2\\
  \angled{m_s \overline{s}s} = -\frac{1}{2} f_K^2 m_K^2
\end{gather}
where PDG values are used for the meson masses~\cite{OliveEtAl2014} 
and the decay constants are~\cite{RosnerStoneVandeRuth2015} 
\begin{equation}
  f_{\pi}=92.2\pm3.5\ \mev\ ,\
  f_K=110.0\pm4.2\ \mev
  .
\end{equation}
The quark mass ratios of strange to light and charm to strange quarks are 
given in~\cite{OliveEtAl2014}; however, in order to consider the RG-invariant product~(\ref{RGinvarQQ}) for all open-flavor combinations of interest, we must combine results from~\cite{OliveEtAl2014} with bottom-flavoured ratios obtained on the lattice~\cite{Chakraborty2015}. 
The resulting ratios and their errors (treated in quadrature) are
\begin{align}
  \frac{M_c}{m_{n}} &= \left( \frac{M_{c}}{m_{s}} \right)\left( \frac{m_{s}}{m_{n}} \right) = 322.6\pm 13.6, & \frac{M_c}{m_{s}} &= 11.73 \pm 0.25,\\
  \frac{M_b}{m_{n}} &= \left( \frac{M_{b}}{M_{c}} \right)\left( \frac{M_{c}}{m_{n}} \right) = 1460.7\pm 64.0, & \frac{M_b}{m_{s}} &= 52.55 \pm 1.30.
\end{align}

For the purely gluonic condensates~(\ref{condensate_gluon_four}) 
and~(\ref{condensate_gluon_six}), 
we use values from~\cite{LaunerNarisonTarrach1984,Narison2010}:
\begin{gather}
  \gluonfourd = (0.075 \pm 0.020)\ \gev^4\label{gluonfourdvalue}\\
  \gluonsixd = \left((8.2 \pm 1.0)\ \gev^2\right)\gluonfourd.
\end{gather}
The 5d mixed condensate can be related to the 3d quark condensate 
through~\cite{LatorreNarisonPascualEtAl1984,Narison1988,Narison2005,Narison2015}
\begin{equation}
\frac{\mixed}{\quarkthreed} \equiv M^{2}_{0} = (0.8\pm 0.1)\,\gev^2.
\end{equation}
Because we are using \eqref{RGinvarQQ} to specify the chiral-violating condensates, in the analysis below, the $\quarkthreed$ effects are subsumed within dimension-four contributions and $\mixed$ effects within dimension-six contributions. 
As noted above, we choose the central value of the renormalization scale $\mu$ to be $M_c$ for the charm systems and $M_b$ for the bottom systems.

\section{Analysis Methodology and Results}\label{IV}
\begin{figure}[htb!]
\centering
\subfloat[]{%
    \includegraphics[width=.8\textwidth]{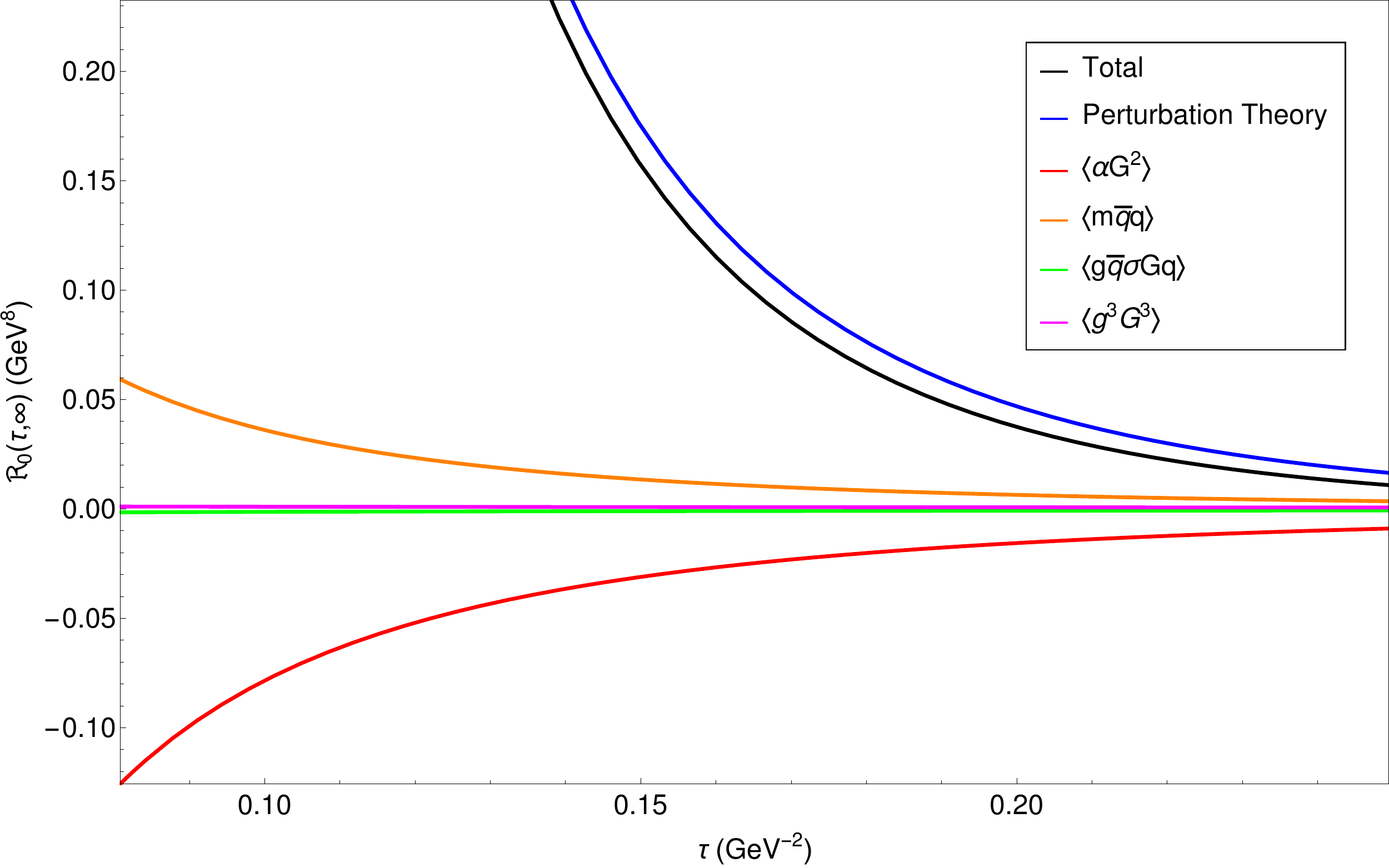}}\\
  \subfloat[]{%
    \includegraphics[width=.8\textwidth]{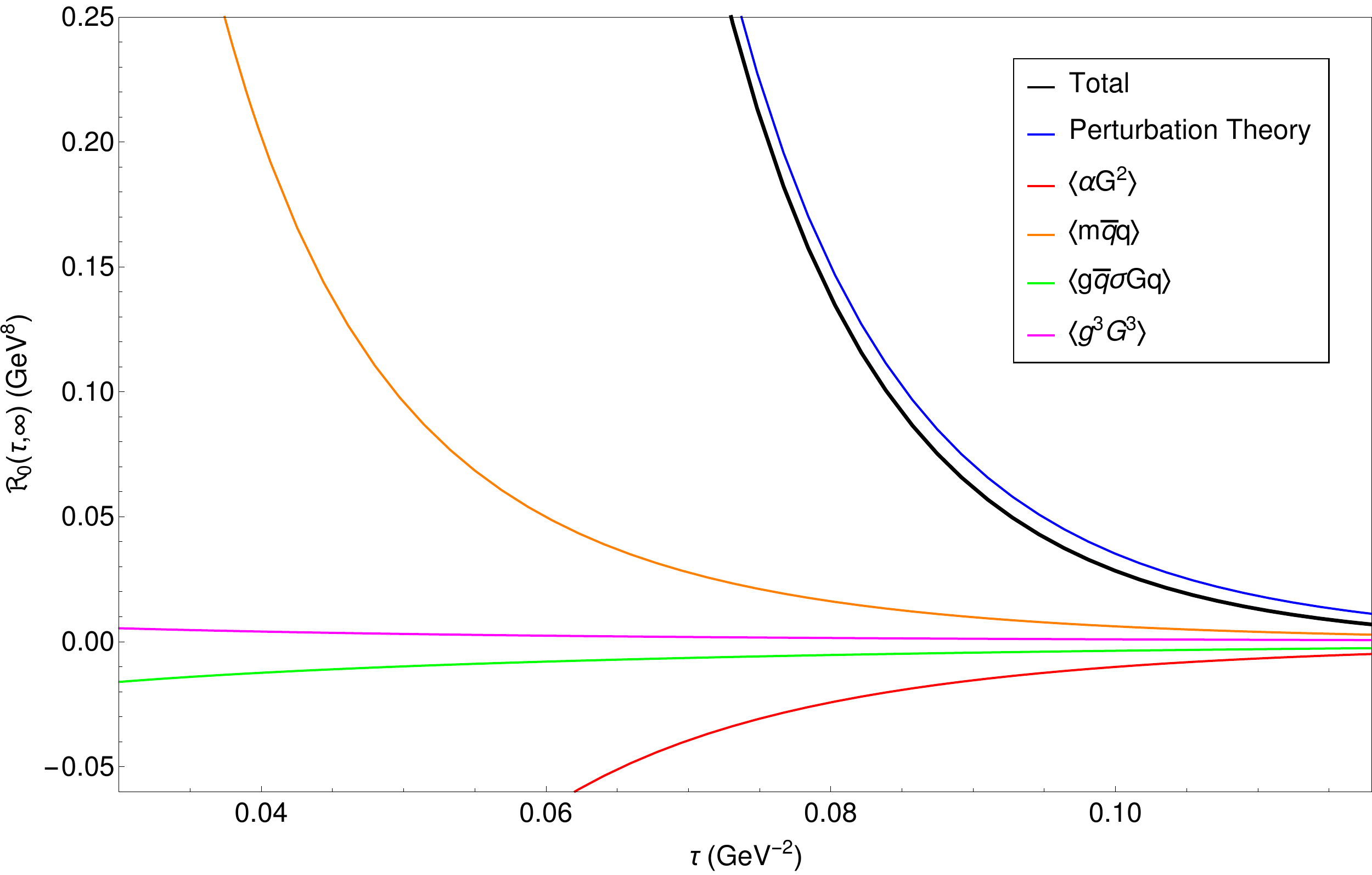}}
\caption{OPE convergence for $0^{+(+)}$ charm-nonstrange and bottom-nonstrange channels.\label{fig.OPEConverge}}
\end{figure}

\begin{figure}[ht!]
\centering
\subfloat[]{%
    \includegraphics[width=0.8\textwidth]{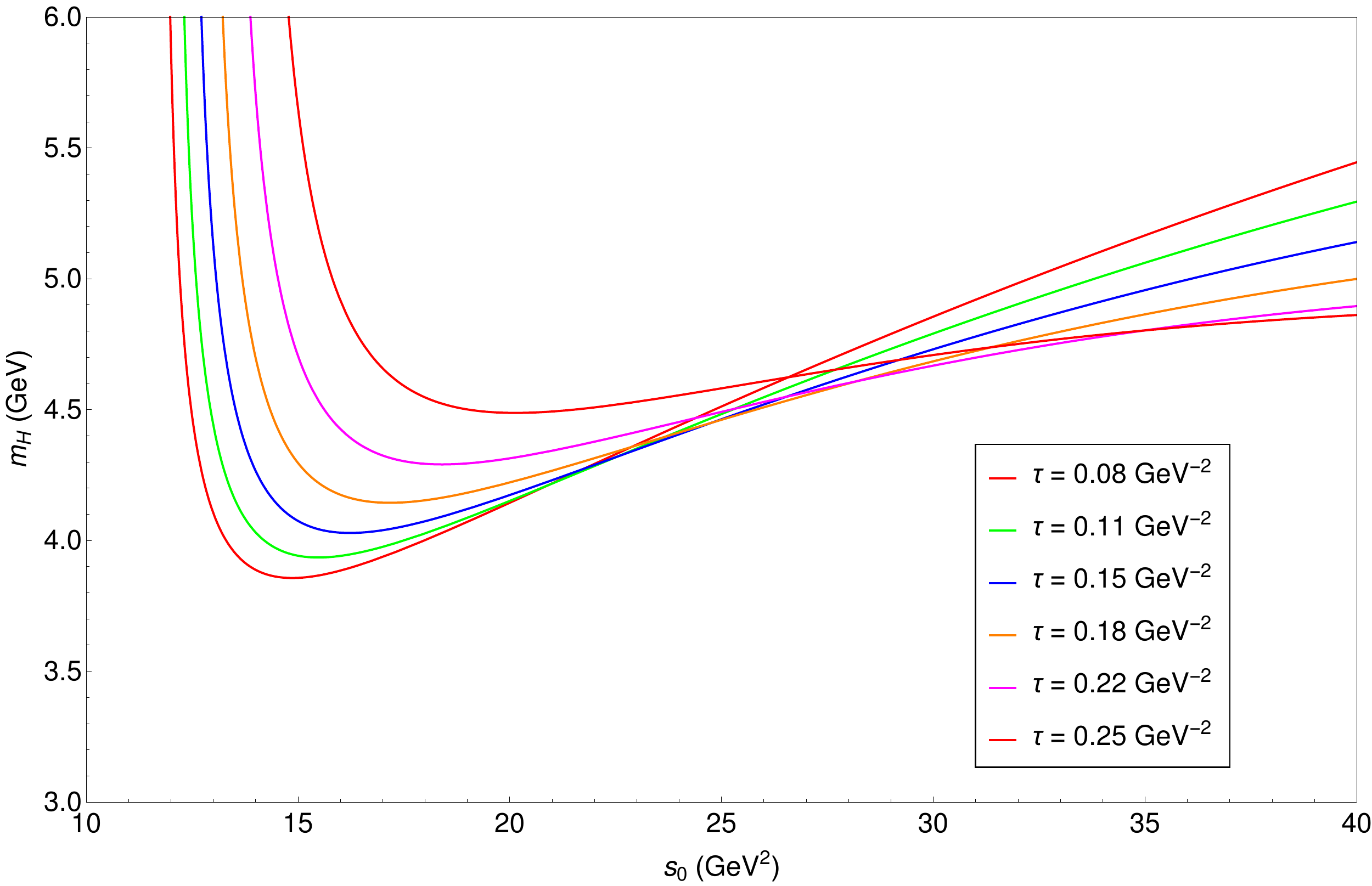}}\\
  \subfloat[]{%
    \includegraphics[width=0.8\textwidth]{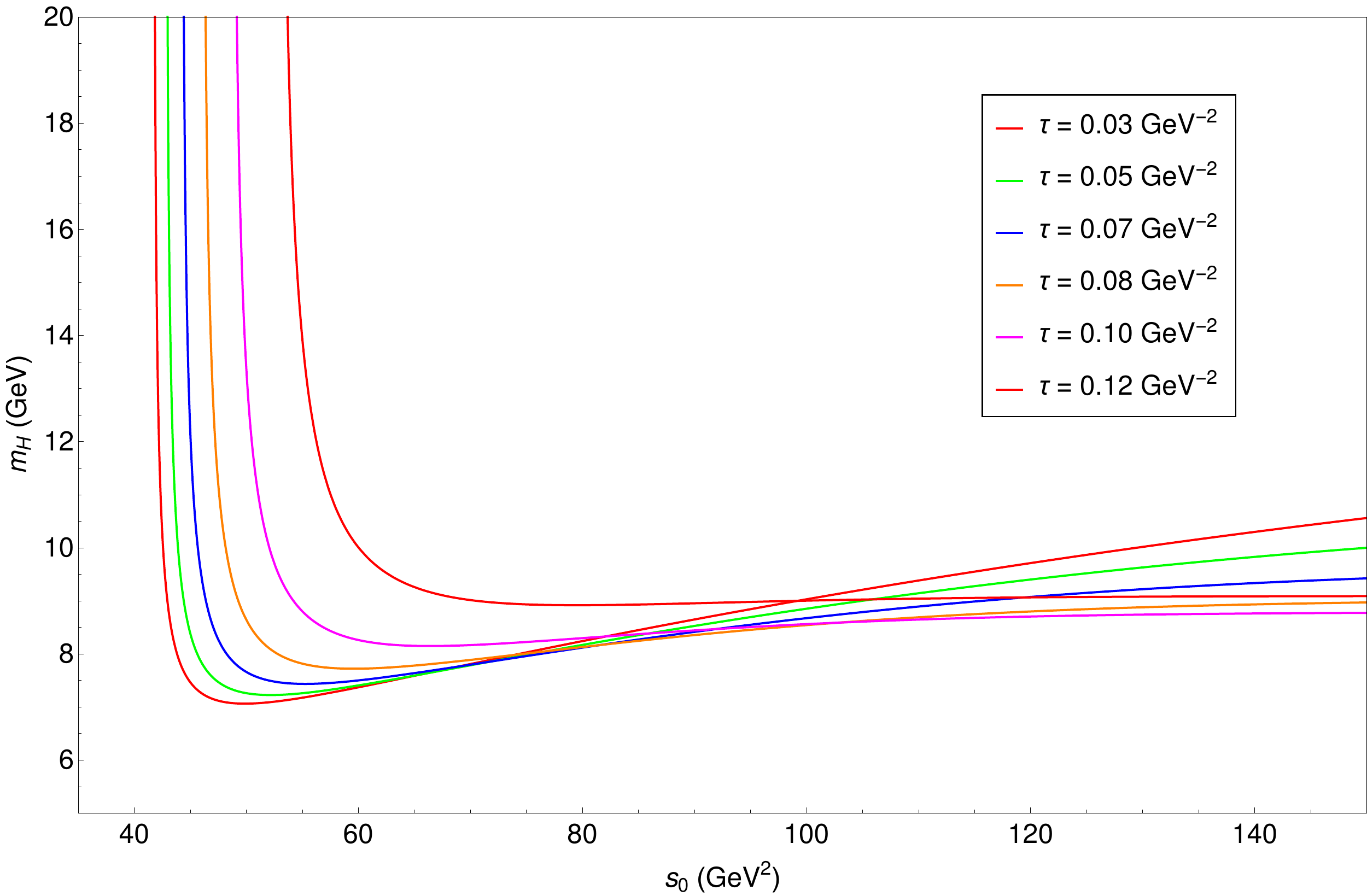}}
\caption{Plots of hybrid mass $m_H$ as a function of continuum threshold $s_0$
for various values of the Borel parameter $\tau$ for $0^{+(+)}$ charm-nonstrange and bottom-nonstrange channels. \label{fig.SumRule}}
\end{figure}

In order to extract stable mass predictions from the QCD sum-rule, we require a suitable range of values for our Borel scale ($\tau$) within which our analysis can be considered reliable. Within this range, we perform a fitting (i.e., minimization) procedure to obtain an optimized value of the continuum onset ($s_0$) associated with our resulting mass prediction.
We determine the bounds of our Borel scale by examining two conditions: the convergence of the OPE, and the pole contribution to the overall mass prediction, mirroring our previous work done in charmonium and bottomonium systems \cite{ChenKleivSteeleEtAl2013}. To enforce OPE convergence and obtain an upper-bound on our Borel window ($\tau_{max}$), we require that contributions to the dimension-four condensate be less than one-third that of the perturbative contribution, and the dimension-six gluon condensate contribute less than one-third of the dimension-four condensate contributions.
(See Figure~\ref{fig.OPEConverge} for an example.)
To determine a lower bound for our Borel window, we examine the pole contribution defined as
\begin{equation}
\mathrm{PC}\left(s_{0}, \tau\right) = \frac{\int_{M_{Q}^{2}}^{s_{0}} e^{-t \tau} \mathrm{Im}\Pi(t) \mathrm{d}t}{\int_{M_{Q}^{2}}^{\infty} e^{-t \tau} \mathrm{Im}\Pi(t) \mathrm{d}t}.
\label{polecontribution}
\end{equation}
The pole contribution constraint can also be understood as a suppression of excited state contributions. To extract a lower bound for our Borel window ($\tau_{min}$), we must first provide a reasonable estimate of the continuum $s_0$ as a seed value for the minimization. To do this, we look for stability in the hadronic mass prediction as a function of $s_{0}$ with variations in $\tau$ (Figure \ref{fig.SumRule}). 
We optimize the initial $s_{0}$ and $m_H$ predictions by minimizing
\begin{equation}
\chi^2 = \sum_{i=1}^{20}\left(\frac{1}{m_{H}}\sqrt{\frac{\lsr_1(\tau_{i},\,s_0 )}{\lsr_0(\tau_{i},\,s_0 )}}-1 \right)^{2},
\label{s0opt}
\end{equation}
where we sum over an equally-spaced discretized $\tau$ range inside the Borel window. 

Minimizing (\ref{s0opt}) results in an optimized values for the continuum  $s_{0}$.  Once $s_{0}$ is found, we may use (\ref{polecontribution}) to determine a lower bound on $\tau$ by requiring a pole contribution of at least $10\%$. Note that this procedure involving (\ref{polecontribution}) should be iterated to ensure that the values of $s_{0}$ and $\tau_{min}$ are self-consistent. 
Once the hadronic mass prediction has been extracted, we may return to (\ref{lsr_main_subtracted}) to solve for the hybrid coupling from the $\tau$ critical point of $f_H$ using the optimized continuum value and hadronic mass prediction. 
\begin{figure}[ht!]
\centering
\subfloat[]{%
    \includegraphics[width=0.8\textwidth]{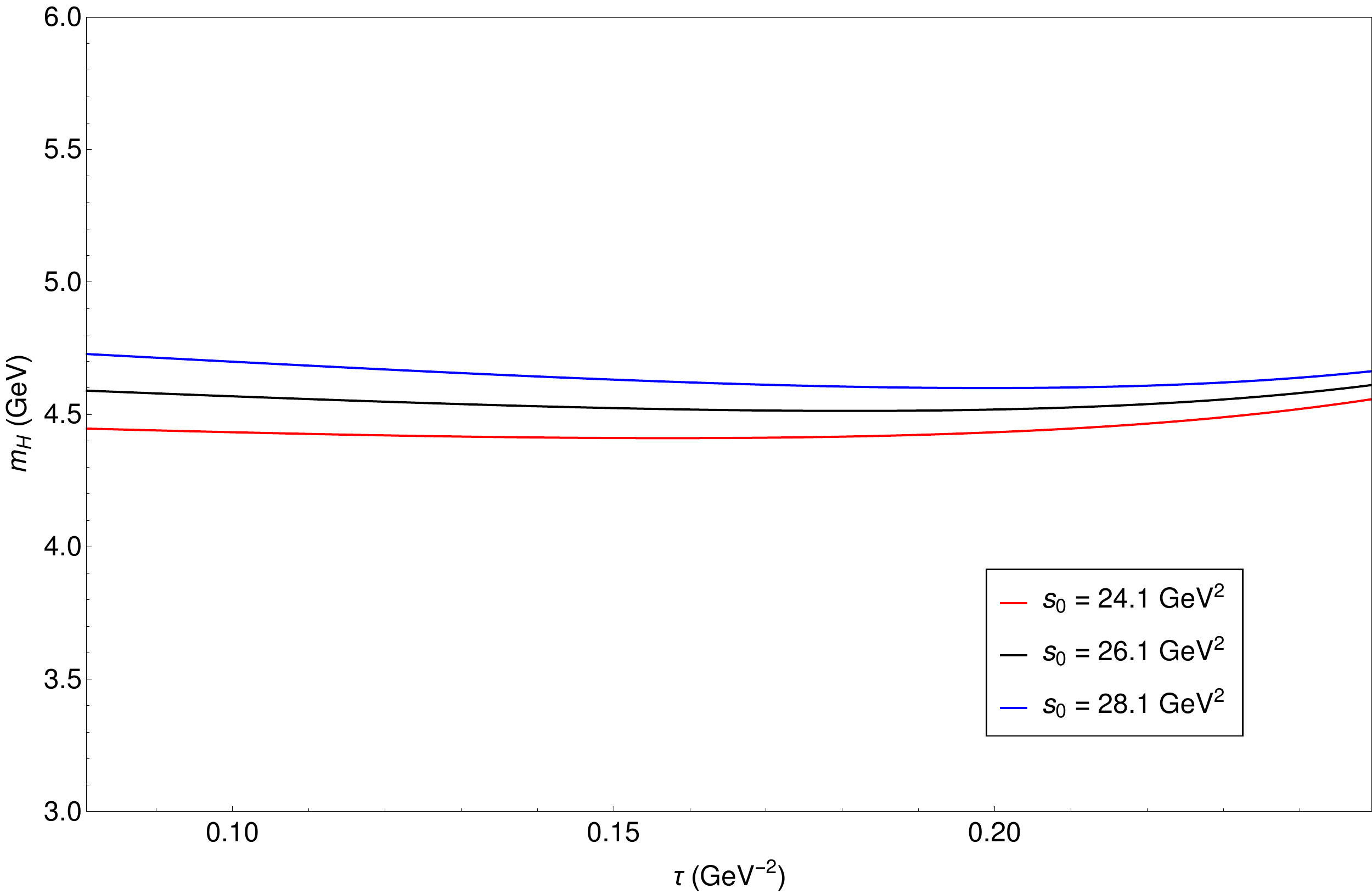}}\\
  \subfloat[]{%
    \includegraphics[width=0.8\textwidth]{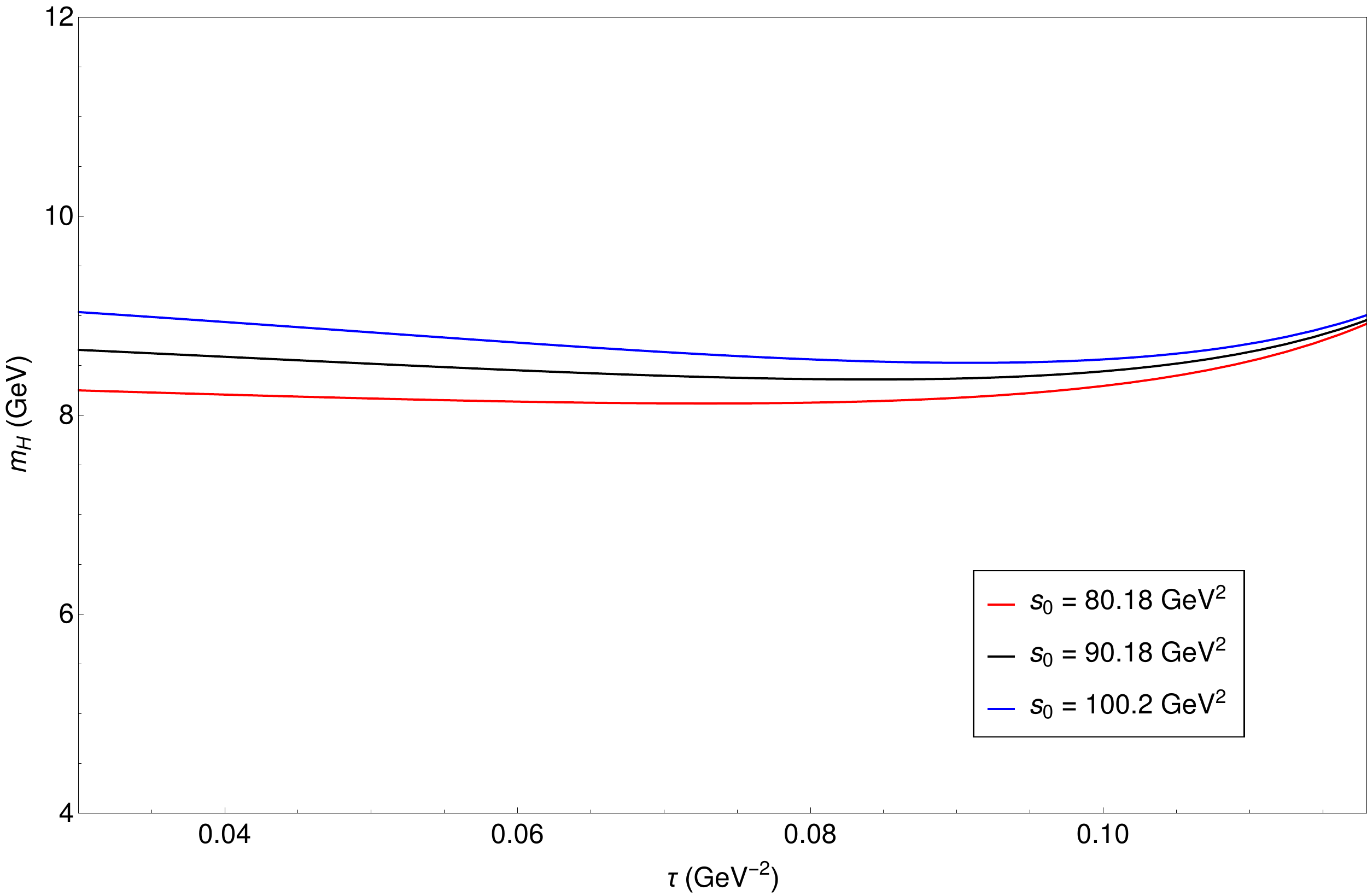}}
\caption{Plots illustrating the stability of mass predictions for $0^{+(+)}$ charm-nonstrange and bottom-nonstrange channels.
\label{fig.Mass}
}
\end{figure}

\begin{figure}[ht]
\centering
\includegraphics[width=0.8\textwidth]{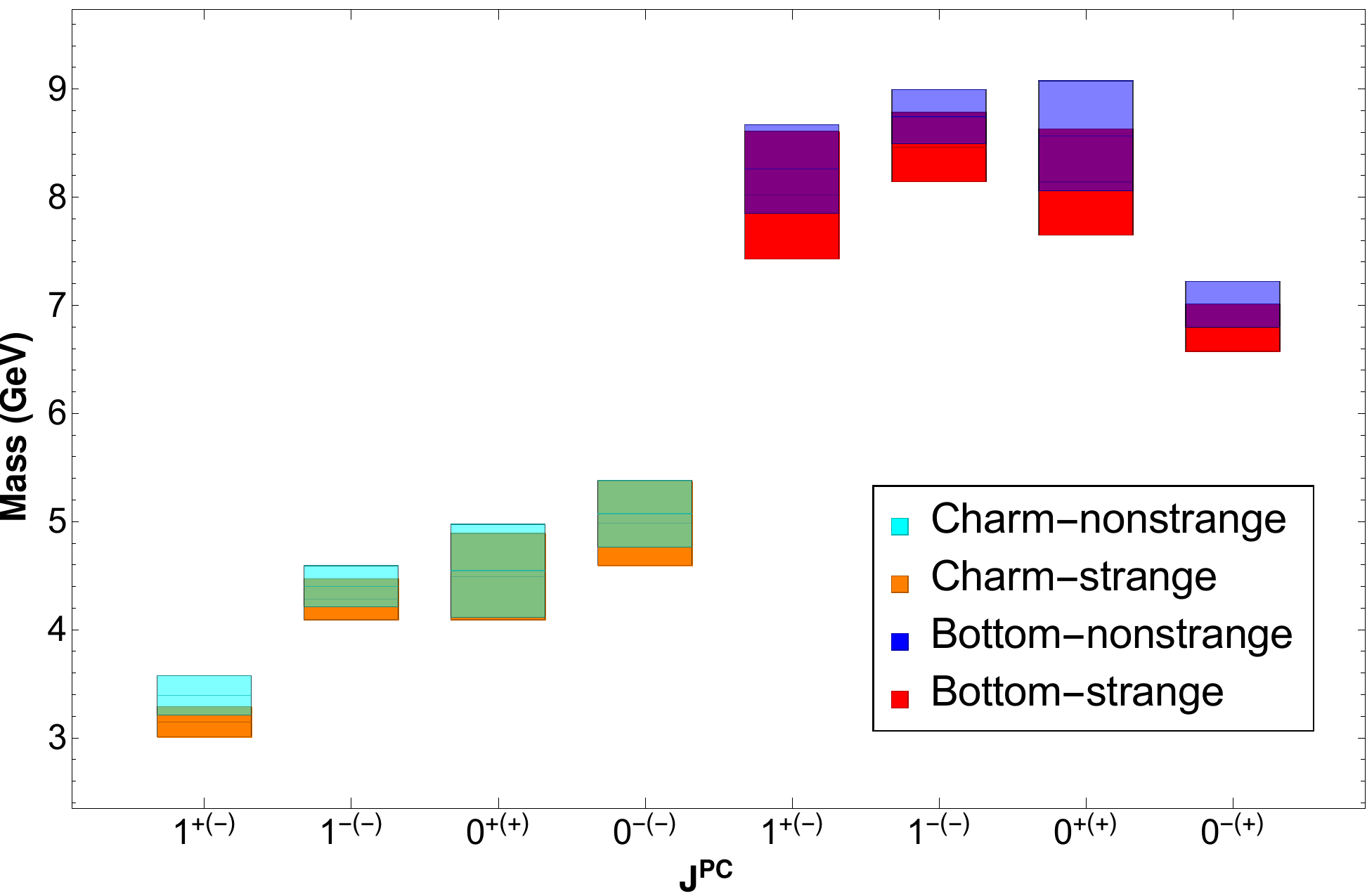}
\caption{Summary of mass predictions in charm and bottom systems with uncertainties.
The overlap between the heavy-nonstrange and heavy-strange predictions is denoted by the green tone in the charm sector and purple in the bottom sector. 
\label{fig.MassSpectra}}
\end{figure}

We present results for the Borel window, continuum, and predicted hybrid mass and couplings
for open-charm and open-bottom hybrids in Tables \ref{D0_results_table} through \ref{Bs_results_table}, and in Figs.~\ref{fig.Mass} and \ref{fig.MassSpectra}. Channels that do not stabilize have been omitted from the tables.  The errors presented encapsulate contributions added in quadrature from the heavy quark masses, quark mass ratios, $\alpha_s$ reference values, and the condensate values. We also include estimations of the error due to truncation of the OPE series by comparing mass predictions with and without 6d contributions and due to variations in the $\tau$ window of $10\%$. Uncertainties associated with the renormalization scale follow the methodology established in 
Ref.~\cite{Narison_RenormScale2015} which doubled the resulting uncertainty associated with variations in the renormalization scale of $\delta\mu=\pm 0.1\,{\rm GeV}$ (charm systems) and $\delta\mu=\pm 0.5\,{\rm GeV}$ (bottom systems).

\begin{table}[!ht]
\centering
\caption{QCD sum-rules analysis results for ground state charm-nonstrange hybrids. }
\label{D0_results_table}
\begin{tabular}{cccccc}
    $J^{PC}$ & $\tau_{\min}\ (\gev^{-2})$ & $\tau_{\max}\ (\gev^{-2})$ 
  & $s_0\pm\delta s_0\ (\gev^2)$ & $m_H\pm\delta m_{H}\ (\gev)$ & $f_{H}^{2}\times10^{6}$\\ 
\hline
  $0^{+(+)}$ & 0.08 & 0.25 & $26.1\pm 6.0$ & $ 4.55\pm 0.43$ & $7.47$\\
  $0^{-(-)}$ & 0.07 & 0.17 & $31.8\pm 4.2$ & $ 5.07\pm 0.31$ & $7.28$\\
  $1^{-(-)}$ & 0.09 & 0.29 & $24.7\pm 2.5$ & $ 4.40\pm 0.19$ & $12.4$\\
  $1^{+(-)}$ & 0.15 & 0.35 & $14.7\pm 1.6$ & $ 3.39\pm 0.18$ & $9.87$
\end{tabular}
\end{table}
\begin{table}[!ht]
\centering
\caption{QCD sum-rules analysis results for ground state charm-strange hybrids. }
\label{Ds_results_table}
\begin{tabular}{cccccc}
    $J^{PC}$ & $\tau_{\min}\ (\gev^{-2})$ & $\tau_{\max}\ (\gev^{-2})$ 
  & $s_0\pm\delta s_0\ (\gev^2)$ & $m_H\pm\delta m_{H}\ (\gev)$ & $f_{H}^{2}\times10^{6}$\\ 
\hline
  $0^{+(+)}$ & 0.08 & 0.24 & $ 25.2\pm 5.2 $ & $ 4.49\pm 0.40$ & $7.36 $\\
  $0^{-(-)}$ & 0.07 & 0.17 & $ 30.4\pm 5.2 $ & $ 4.98\pm 0.39$ & $2.03 $\\
  $1^{-(-)}$ & 0.10 & 0.30 & $ 23.1\pm 2.4 $ & $ 4.28\pm 0.19$ & $11.0 $\\
  $1^{+(-)}$ & 0.18 & 0.34 & $ 12.5\pm 1.1 $ & $ 3.15\pm 0.14$ & $8.45 $
\end{tabular}
\end{table}
\begin{table}[!ht]
\centering
\caption{QCD sum-rules analysis results for ground state bottom-nonstrange hybrids. }
\label{B0_results_table}
\begin{tabular}{cccccc}
    $J^{PC}$ & $\tau_{\min}\ (\gev^{-2})$ & $\tau_{\max}\ (\gev^{-2})$ 
  & $s_0\pm\delta s_0\ (\gev^2)$ & $m_H\pm\delta m_{H}\ (\gev)$& $f_{H}^{2}\times10^{6}$\\ 
\hline
  $0^{+(+)}$ & 0.03 & 0.12 & $92.5\pm 15.6$ & $8.57\pm 0.51$ & $1.28$ \\
  $0^{-(+)}$ & 0.05 & 0.09 & $59.1\pm 3.9$ & $7.01 \pm 0.21$ &  $0.516$\\
  $1^{-(-)}$ & 0.03 & 0.10 & $94.7\pm 7.5$ & $8.74\pm 0.25$ & $1.76$\\
  $1^{+(-)}$ & 0.03 & 0.14 & $86.7\pm 11.1$ & $8.26\pm 0.41$ & $1.66$
\end{tabular}
\end{table}
\begin{table}
\centering
\caption{QCD sum-rules analysis results for ground state bottom-strange hybrids. }
\label{Bs_results_table}
\begin{tabular}{cccccc}
    $J^{PC}$ & $\tau_{\min}\ (\gev^{-2})$ & $\tau_{\max}\ (\gev^{-2})$ 
  & $s_0\pm\delta s_0\ (\gev^2)$ & $m_H\pm\delta m_{H}\ (\gev)$& $f_{H}^{2}\times10^{6}$\\ 
\hline
  $0^{+(+)}$ & 0.04 & 0.11 & $79.9\pm 13.0$ & $8.14\pm 0.49$ &$0.817$\\
  $0^{-(+)}$ & 0.06 & 0.10 & $55.1\pm 4.0$ & $6.79\pm 0.22$ &$0.434$\\
  $1^{-(-)}$ & 0.03 & 0.10 & $87.6\pm 9.9$ & $8.46\pm 0.32$ &$1.24$ \\
  $1^{+(-)}$ & 0.04 & 0.15 & $81.7\pm 15.7$ & $8.02\pm 0.59$ &$1.39 $
\end{tabular}
\end{table}

%%%%%%%%%%%%%%%%%%%%%%%%%%%%%%%%%%%%%%%%%%%%%%%%%%%%%%%%%%%%%%%%%%%%%%%%%%%%%%%%%%%%%%%%%%%%%%%%
%
%
%%%%%%%%%%%%%%%%%%%%%%%%%%%%%%%%%%%%%%%%%%%%%%%%%%%%%%%%%%%%%%%%%%%%%%%%%%%%%%%%%%%%%%%%%%%%%%%%

\begin{table}
\centering
\caption{Charm-Nonstrange Masses from Higher-Weight Sum-Rules (GeV)}
\label{charmnonstrange-higherSR}
\begin{tabular}{llll}
 $J^{P(C)}$& $\sqrt{\frac{R_{1}}{R_{0}}}$ & $\sqrt{\frac{R_{2}}{R_{1}}}$ & $\sqrt{\frac{R_{3}}{R_{2}}}$ \\
 \hline
 $0^{+(+)}$& 4.54 & 4.54 & 4.59 \\
 $0^{-(-)}$& 5.07 & 5.07 & 5.12 \\
 $1^{-(-)}$& 4.40 & 4.39 & 4.45 \\
 $1^{+(-)}$& 3.39 & 3.39 & 3.45
\end{tabular}
\end{table}
\begin{table}
\centering
\caption{Bottom-Nonstrange Masses from Higher-Weight Sum-Rules (GeV)}
\label{bottomnonstrange-higherSR}
\begin{tabular}{llll}
 $J^{P(C)}$& $\sqrt{\frac{R_{1}}{R_{0}}}$ & $\sqrt{\frac{R_{2}}{R_{1}}}$ & $\sqrt{\frac{R_{3}}{R_{2}}}$ \\
 \hline
 $0^{+(+)}$& 8.57 & 8.52 & 8.60 \\
 $0^{-(+)}$& 7.01 & 7.01 & 7.06 \\
 $1^{-(-)}$& 8.74 & 8.71 & 8.80 \\
 $1^{+(-)}$& 8.26 & 8.20 & 8.29 \\
\end{tabular}
\end{table}
\begin{figure}[ht!]
\centering
\subfloat[]{%
    \includegraphics[width=0.8\textwidth]{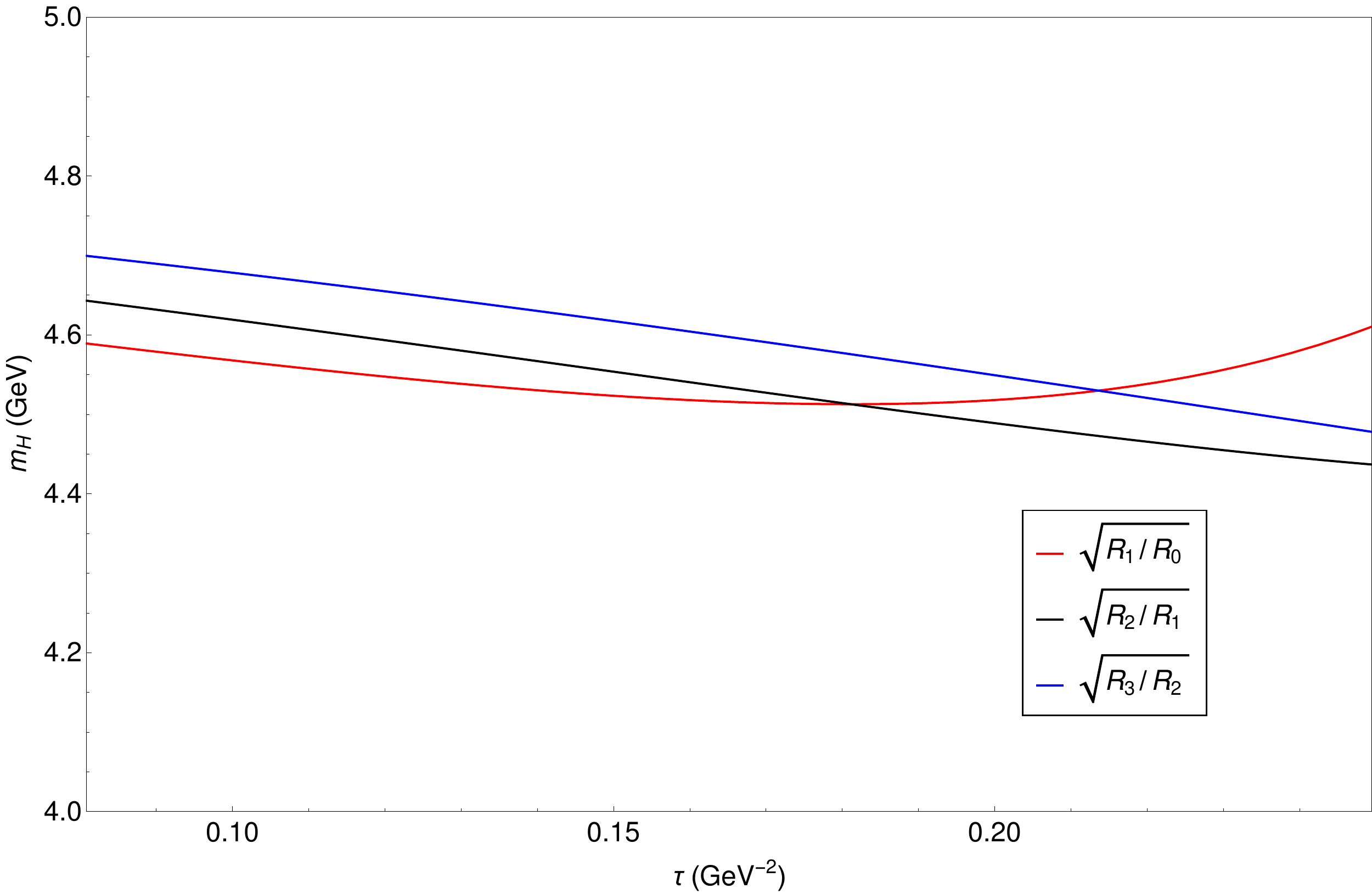}}\\
  \subfloat[]{%
    \includegraphics[width=0.8\textwidth]{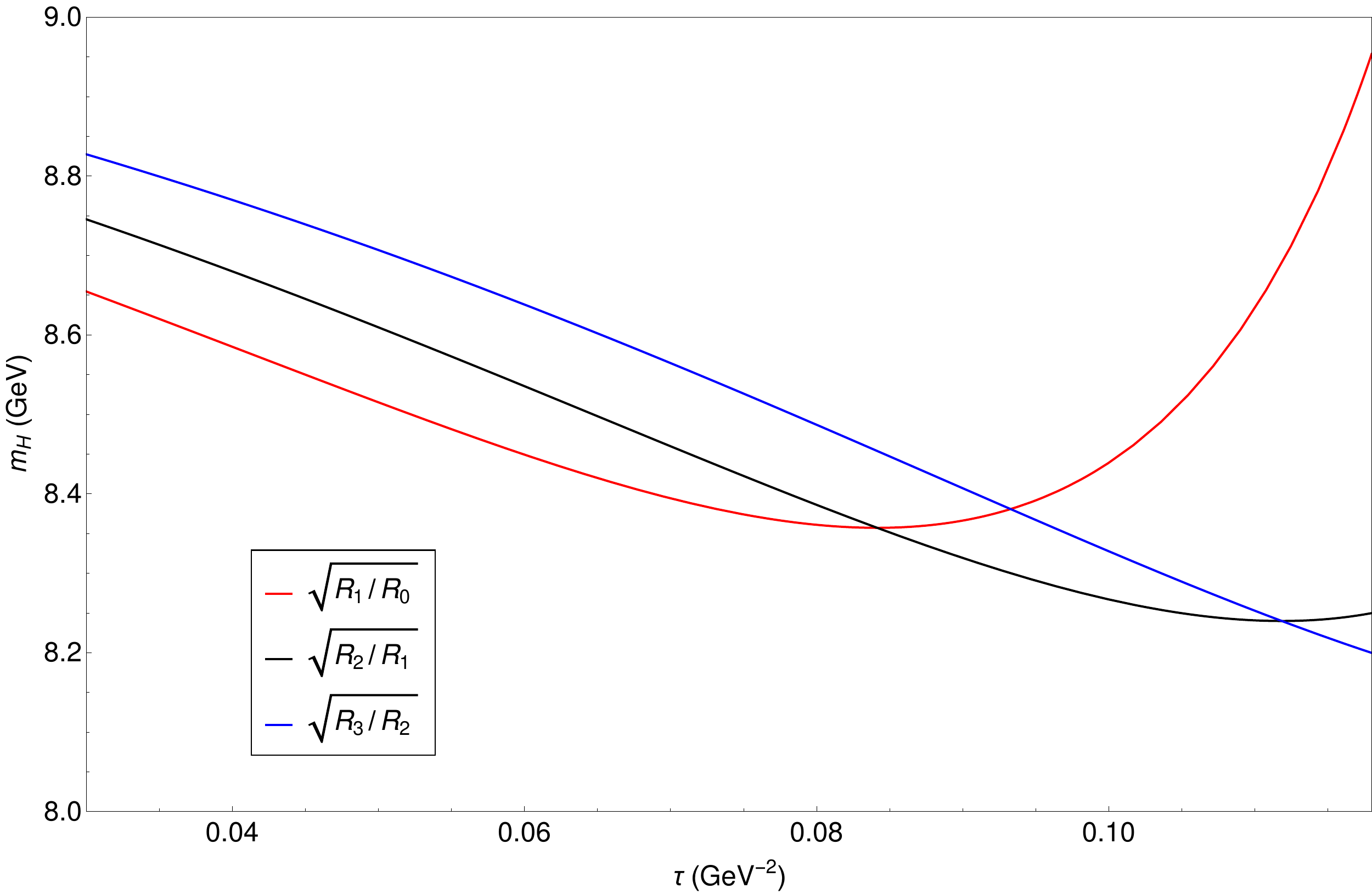}}
\caption{Plots illustrating higher-weight sum-rule ratios in $0^{+(+)}$ charm-nonstrange and bottom-nonstrange channels.
\label{fig.HigherWeightSR}
}
\end{figure}
As a validation of our analysis, we also consider ratios of higher-weight sum-rules which serve as a generalization of \eqref{lsr_master}: 
\begin{equation}\label{lsr_hw_ratio}
  \frac{\lsr_{k+1}(\tau,\,s_0 )}{\lsr_k(\tau,\,s_0 )}=m_H^2 ~.
\end{equation}
In Table \ref{charmnonstrange-higherSR}, Table \ref{bottomnonstrange-higherSR} and Figure \ref{fig.HigherWeightSR}  
we compare the nonstrange sum-rule ratios for $k=0,1,2$.  Although the higher-weight ratios have greater sensitivity to the high-energy region of the spectral function (excited states and QCD continuum), the hadronic mass scales emerging from the various weights are remarkably consistent, indicating that the sum-rule window has been well-chosen to emphasize the lightest hybrid state via the pole contribution criterion \eqref{polecontribution}.

%%%%% begin moved stuff

\section{Mixing Effects}\label{new_section}
\begin{figure}[ht]
\centering
\subfloat[]{%
    \includegraphics[width=0.8\textwidth]{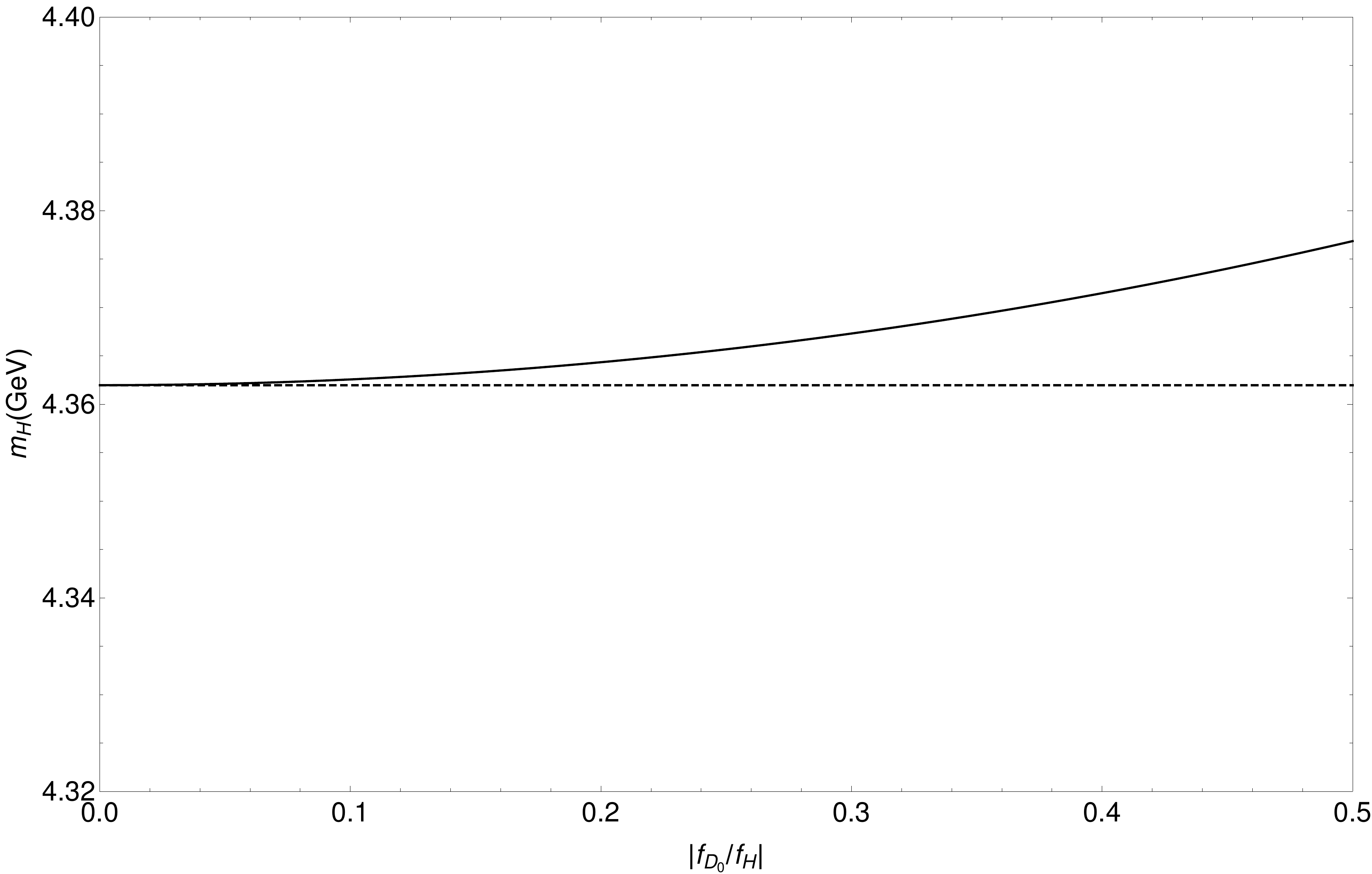}\label{fig.MixingD0}}\\
  \subfloat[]{%
    \includegraphics[width=0.8\textwidth]{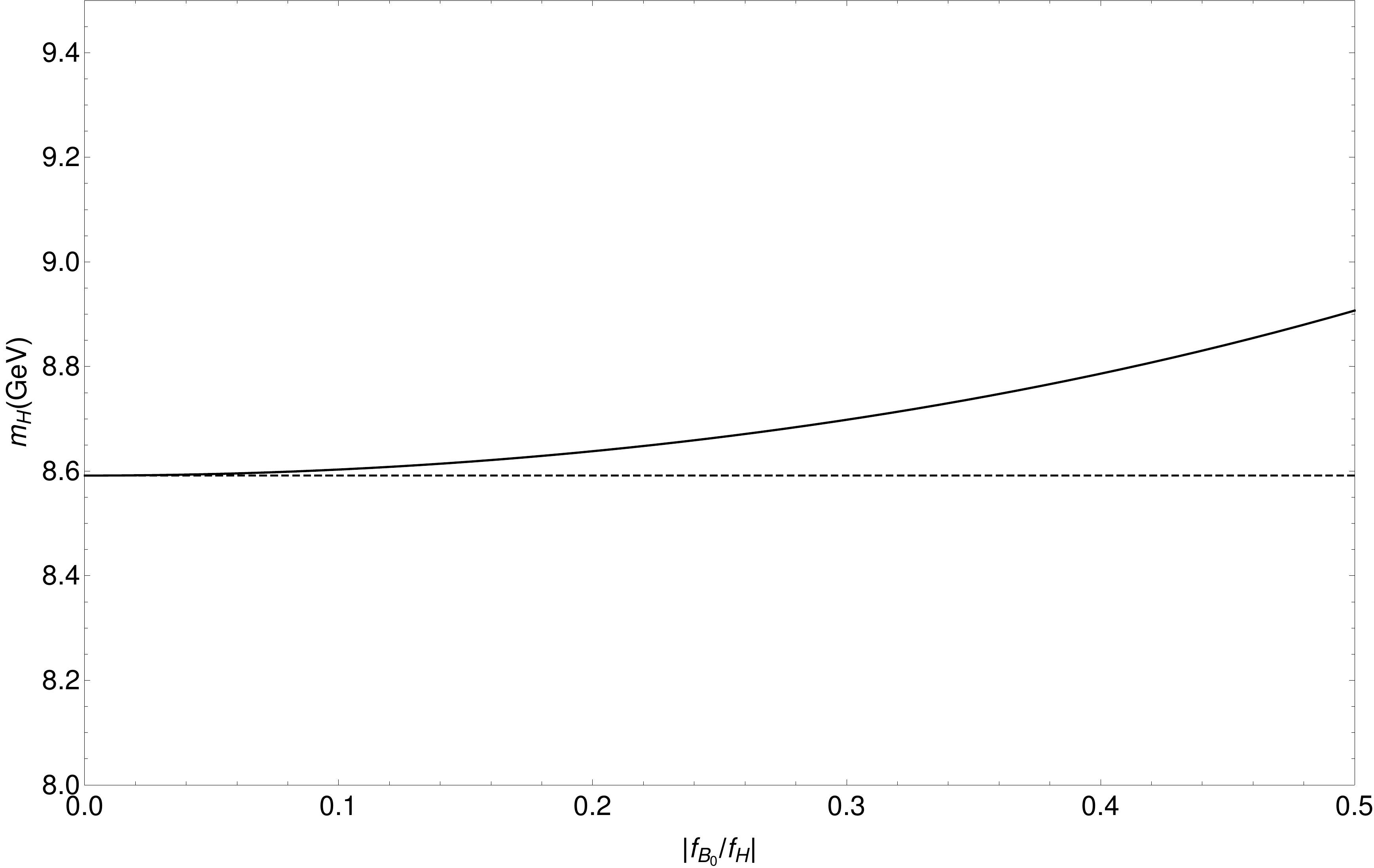}\label{fig.MixingB0}}
\caption{Dependence of hybrid mass $m_{H}$ defined in (\ref{lsr_master_mixed}) on the conventional state coupling constant $f_{conv}$ for $1^{-(-)}$ \protect\subref{fig.MixingD0} charm-nonstrange and \protect\subref{fig.MixingB0} bottom-nonstrange states taken at the central $\tau$ values from Tables \ref{D0_results_table} and \ref{B0_results_table}. Dashed lines indicate the mass prediction associated with the channel. An upper bound has been placed on the mixing effect by considering that the  coupling of conventional states to the hybrid current is restricted to $\left\vert\frac{f_{conv}}{f_H}\right\vert \leq \frac{1}{2}$.
\label{fig.Mixing}}
\end{figure}

As noted in Section~\ref{I}, the open-flavor structure of the hybrid systems in question precludes 
the possibility of explicitly exotic $J^{PC}$ states. 
As such, we might expect a degree of mixing with conventional mesonic states. 
In our previous work on heavy quarkonium hybrids~\cite{ChenKleivSteeleEtAl2013}, 
this possibility of mixing was examined 
through the addition of a conventional meson to the single narrow resonance model (\ref{single_narrow_resonance}) such that~(\ref{lsr_main_subtracted}) becomes
\begin{equation}\label{lsr_main_subtracted_mixed}
  \lsr_k(\tau,\,s_0 ) = f_H^2 m_H^{8+2k} e^{-m_H^2 \tau} + f_{conv}^2 m_{conv}^{8+2k} e^{-m_{conv}^2 \tau}
\end{equation}
where the parameters $f_{conv}$ and $m_{conv}$ are the coupling constant and mass of the ground state conventional meson sharing the same $J^P$ values. 
By including these terms, we can form a sum-rule coupled to the conventional state,
\begin{equation}\label{lsr_master_mixed}
m_H^2=\frac{\lsr_1(\tau,\,s_0 )-f_{conv}^2 m_{conv}^{10} e^{-m_{conv}^2 \tau}}{\lsr_0(\tau,\,s_0 )-f_{conv}^2 m_{conv}^{8} e^{-m_{conv}^2 \tau}}
\end{equation}
which can be used to investigate the dependence of the hybrid mass on the coupling to the conventional state by using known values of conventional meson masses to specify $m_{conv}$. We see in the resulting Figure \ref{fig.Mixing} that increasing the coupling to the conventional state tends to increase the hybrid mass prediction, indicating our results presented here may correspond to a lower bound on the hybrid mass if mixing with conventional states is substantial. From Figure \ref{fig.Mixing}, we estimate an upper bound on the increased hybrid mass by implementing the condition that the coupling of the hybrid current to the conventional state $f_{conv}$ be no more than half the coupling of the hybrid current to the hybrid state $f_{H}$ (Tables \ref{D0_results_table} to \ref{Bs_results_table}). In the simplest mixing scenario this limit on $f_{conv}$ corresponds to a mixing angle of approximately half that of a maximal mixing between conventional and hybrid mesons.
The estimated effect of mixing on the hybrid mass prediction is summarized in Table \ref{mixing_error}, and shows interesting dependence on $J^P$.

\begin{table}
\centering
\caption{Effect on hybrid mass prediction from mixing with conventional meson states. Masses from Tables \ref{D0_results_table} to \ref{Bs_results_table} are summarized with $\delta m_{mix}$ expressing increased mass range with mixing up to $\left\vert\frac{f_{conv}}{f_H}\right\vert = \frac{1}{2}$ due to coupling to the lowest-lying conventional state with appropriate quantum numbers according to PDG \cite{OliveEtAl2014}. Entries have been omitted where no conventional meson state has been tabulated.
} 
\label{mixing_error}
\begin{tabular}{cccccc}
    Flavour & $J^{P}$ & $m_{H}~(\gev)$ & PDG State & $m_{conv} (\gev)$ & $+\delta m_{mix}~(\gev)$ \\ 
\hline
 Charm-nonstrange & $0^{+}$ & $4.54$ &  $D^{*}_{0}\left(2400\right)^0$ & $2.318$ & 0.02 \\
  				  & $0^{-}$ & $5.07$ &  $D^{0}$ & $1.865$ & 0.00 \\
  				  & $1^{-}$ & $4.40$ &  $D^{*}\left(2007\right)^0$ & $2.007$ & 0.01 \\
    			  & $1^{+}$ & $3.39$ &  $D_{1}\left(2420\right)^0$ & $2.420$ & 0.05 \\
 Charm-strange    & $0^{+}$ & $4.49$ &  $D^{*}_{s0}\left(2317\right)^{\pm}$ & $2.318$ & 0.02 \\
  				  & $0^{-}$ & $4.98$ &  $D^{\pm}_{s}$ & $1.969$ & 0.00 \\
  				  & $1^{-}$ & $4.28$ &  $D^{*\pm}_{s}$ & $2.112$ & 0.02 \\
    			  & $1^{+}$ & $3.15$ &  $D_{s1}\left(2460\right)^{\pm}$ & $2.460$ & 0.06 \\
 Bottom-nonstrange & $0^{+}$ & $8.57$ & - & - & -\\
  				  & $0^{-}$ & $7.01$ &  $B_{0}$ & $5.279$ & 0.19 \\
  				  & $1^{-}$ & $8.74$ &  $B^{*}$ & $5.324$ & 0.32 \\
    			  & $1^{+}$ & $8.26$ &  $B_{1}\left(5721\right)^{0}$ & $5.726$  & 0.74 \\
 Bottom-strange   & $0^{+}$ & $8.14$ &  - & - & -\\
  				  & $0^{-}$ & $6.79$ &  $B^{0}_{s}$ & $5.367$ & 0.44 \\
  				  & $1^{-}$ & $8.46$ &  $B^{*}_{s}$ & $5.416$ & 0.35 \\
    			  & $1^{+}$ & $8.02$ &  $B_{s1}\left(5830\right)^0$ & $5.828$ & 0.72 
\end{tabular}
\end{table}

%%%%% end moved stuff

\section{Discussion}\label{V}
For each open-flavour heavy-light hybrid combination under consideration, 
we performed a LSRs analysis
of all eight $J^{P(C)}$ combinations defined according to Table~\ref{JPC_table}.
As can be inferred from Tables~\ref{D0_results_table}--\ref{Bs_results_table} as well as Figure \ref{fig.MassSpectra}, half of the 
analyses stabilized; the other half did not. 
In particular, the $J^{P(C)}\in \{0^{+(+)},1^{-(-)},1^{+(-)}\}$ analyses were stable while the
$J^{P(C)}\in \{0^{+(-)},1^{+(+)},1^{-(+)}\}$ were unstable.
For charm-light hybrids, 
the $0^{-(-)}$ sector stabilized whereas the $0^{-(+)}$ sector did not.
For bottom-light hybrids, this situation was reversed:
the $0^{-(+)}$ sector stabilized while the $0^{-(-)}$ sector did not.
This should be contrasted with GRW for which the stable channels were
$J^{P(C)}\in\{0^{+(+)},0^{-(-)},1^{+(+)},1^{-(-)}\}$ for all heavy-light flavour hybrids.
Comparing to GRW by truncating our additional condensate terms, we find that this change in stability originates from the addition of the 5d and 6d condensate terms.
Note that, for all heavy-light quark combinations considered, 
we did arrive at a unique mass prediction for each $J^P$.
GRW found something similar,
but, as can be seen from Tables~\ref{GRW_CharmCompare} 
and~\ref{GRW_BottomCompare}, the central value of our mass predictions differ significantly from that of GRW in all channels except the $1^{+}$ charm-nonstrange.  However, 
we note that GRW observed a change in the $C$ value for currents that stabilized as the heavy quark mass increased, a feature shared in our analysis where the charm $0^{--}$ and bottom $0^{-+}$ channels stabilized.

%\clearpage
\begin{table}[t]
\centering
\caption{Comparison of central values against GRW mass predictions for $\overline{c}qG$ hybrids ($q=\{u,d\}$).} 
\label{GRW_CharmCompare}
\begin{tabular}{cccccc}
    $J^{P}$ & $m_{\mathrm{GRW}}(\gev)$ & $m_{\mathrm{H}}\ (\gev)$ \\ 
\hline
  $0^{+}$ & $4.0$ &  $ 4.54$ \\
  $0^{-}$ & $4.5$ &  $ 5.07$  \\
  $1^{-}$ & $3.6$ & $ 4.40$  \\
  $1^{+}$ & $3.4$ & $ 3.39$  \\
\end{tabular}
\end{table}

\begin{table}
\centering
\caption{Comparison of central values against GRW mass predictions for $\overline{b}qG$ hybrids ($q=\{u,d\}$).}
\label{GRW_BottomCompare}
\begin{tabular}{cccccc}
    $J^{P}$ & $m_{\mathrm{GRW}}(\gev)$ & $m_{\mathrm{H}}\ (\gev)$ \\ 
\hline
  $0^{+}$ & $6.8$ & $8.57$ \\
  $0^{-}$ & $7.7$ & $7.01$   \\
  $1^{-}$ & $6.7$ & $8.74$  \\
  $1^{+}$ & $6.5$ & $8.26$  \\
\end{tabular}
\end{table}

In all stable channels, the most significant non-perturbative contribution to the LSRs
is the 4d gluon condensate term of the OPE.
At the corresponding optimized value of $s_0$ and over the Borel window indicated in 
Tables~\ref{D0_results_table}--\ref{Bs_results_table}, 
the 4d gluon condensate term accounts for
roughly 10--30\% of the area underneath the $\lsr_0(\tau,s_0)$ curve.
The second most significant contribution comes from the 3d quark condensate term
which accounts for roughly 10\% of the area while
the 5d mixed and 6d gluon condensate contributions each account for $\lesssim 5$\%.
Light quark mass corrections to massless perturbation theory 
are numerically insignificant leading to isospin invariance of our results.

The dominant contributions to the error in both the charm and bottom systems come from the gluon condensates, and the truncation of the OPE. All channels are relatively insensitive to uncertainties in the quark condensate, the heavy quark masses, the quark mass ratios, the reference values of $\alpha_s$, and variations in the $\tau$ range and renormalization scales.

Within computational uncertainty, we cannot preclude degeneracy between the mass spectra of the heavy-nonstrange hybrid systems 
and their heavy-strange counterparts. 
(Compare Tables~\ref{D0_results_table} and~\ref{Ds_results_table}
as well as Tables~\ref{B0_results_table} and~\ref{Bs_results_table}. 
Also, see Figure~\ref{fig.MassSpectra}.)
This can be attributed to the small size of the light quark mass correction to 
massless perturbation theory and to the presence of a heavy quark mass factor
as opposed to a light quark mass factor
in the 3d quark and 5d mixed condensate contributions to the OPE.

Apart from the $0^-$ states, both the charm and bottom cases share a mass hierarchy pattern for the $1^+$, $1^-$ and $0^+$ states where the $1^{+}$ state is lighter than essentially degenerate $1^{-}$ and $0^{+}$ states.  The $0^-$ states have different roles in the mass hierarchies in the charm and bottom sector, which we hypothesize as originating from the differing $C$ quantum numbers associated with their currents. Although open-flavour systems do not have a well-defined $C$ quantum number,  Ref.~\cite{HilgerKrassnigg} attributes physical meaning to $C$ in the internal structures of hybrids and finds that the $0^{-(-)}$  structure is heavier than the $0^{-(+)}$, identical to the pattern we observe in Fig.~\ref{fig.MassSpectra}.

In GRW, for each heavy-light hybrid channel whose LSR analysis was stable, the authors pointed 
out that the difference between the square of the predicted resonance mass and the continuum
threshold parameter was small, typically a couple of hundred MeV which did not seem to allow
for much in the way of resonance width.
In our updated analysis, Tables~\ref{D0_results_table}--\ref{Bs_results_table}
shows that even a relatively wide resonance would be well-separated from the
continuum.

We can compare our negative parity charm hybrid results to those computed 
on the lattice in~\cite{MoirPeardonRyanEtAl2013}. In general, our predictions are heavier and show a larger mass splitting between states. 

% Concluding Paragraph
In summary,
we have performed a QCD LSR analysis of spin-0,1, heavy-light open flavour hybrids. 
In the OPE, we included condensates up to dimension-six as well as 
leading-order light quark mass corrections to massless perturbation theory.
For all flavour combinations, we extracted a single mass prediction for each $J^P\in\{0^{\pm},\,1^{\pm}\}$
(see Tables~\ref{D0_results_table}--\ref{Bs_results_table}).
 Our results were isospin-invariant and within theoretical uncertainties, we could not preclude degeneracy under the exchange of light nonstrange and strange quarks. 
We find similar mass hierarchy patterns in the charm and bottom sectors for the $1^{\pm}$ and $0^+$ states, and that Ref.~\cite{HilgerKrassnigg} provides a natural interpretation for our $0^-$ mass predictions.
Finally, given that open-flavour hybrids cannot take on exotic $J^{PC}$, mixing with
conventional mesons could be important; our analysis suggests that such mixing would tend to 
increase the hybrid mass predictions, and we have estimated an upper bound on this effect.
%%%%%%%%%%%%%%%%%%%%%%%%%%%%%%%%%%%%%%%%%%%%%%%%%%%%%%%%%%%%%%%%%%%%%%%%%%%%%%%%%%%%%%%%%%%%%%%%%%%%%%%%%%%%
\clearpage
\acknowledgments
We are grateful for financial support from the Natural Sciences and 
Engineering Research Council of Canada (NSERC).
DH is thankful for the hospitality provided by the University of
Saskatchewan during his sabbatical.
\bibliographystyle{h-physrev}
\bibliography{research}

\end{document}